\documentclass[twocolumn, showkeys,amsmath,amssymb,nofootinbib,pra]{revtex4}
\usepackage{mathrsfs}
\usepackage{graphicx}
\usepackage{dcolumn}
\usepackage{bm}
\usepackage{physics}
\usepackage{amsmath}
\usepackage{mathtools}
\usepackage{float}
\usepackage[dvips,frame,rotate,import]{xy}
\xyoption{arc}
\usepackage{amsmath}
\DeclareMathOperator{\Sech}{sech}
\DeclareMathOperator{\Tanh}{tanh}
\usepackage{amssymb}
\usepackage{revsymb}
\usepackage{adjustbox}
\usepackage[mathscr]{eucal}
\usepackage{mathptmx}
\usepackage[11pt]{moresize}
\usepackage{tikz}

\begin{document}

\title{Spin-2 BEC spinor superfluid soliton-soliton scattering in one and two space dimensions}

\author{Jasper Taylor, Steven Smith, Jeffrey Yepez}
\date{July 30, 2019}

\address{
Department of Physics and Astronomy, University of  Hawai`i at Manoa\\
Watanabe Hall, 2505 Correa Road, Honolulu, Hawai`i 96822
 }

\begin{abstract}
Presented is a study of  a spin-2 Bose-Einstein condensate (BEC) by unitary quantum simulations of time-dependent soliton-soliton scattering.  The quantum simulation method is based on a  quantum lattice algorithm which is designed for implementation on a future digital quantum computer but is tested  today using a parallel computing architecture based on graphical processing units (GPUs).   
We analytically solve the spin-2 BEC equations of motion, a nonlinear system of 5 coupled Gross-Pitiaevskii (GP) equations, in one- and two-spatial dimensions.  In 1D there are 16 bright soliton and 16 dark soliton solutions.  In 2D there are 3 dark solition solutions Pad\'e approximation solutions, for $m_f=\pm2$, $m_f=\pm 1$ and $m_f=0$,  corresponding to quantum vortices. 
We report on the implementation the unitary quantum lattice gas algorithm for spinor superfluid and establish its efficacy by validating the stability of the 1D and 2D energy eigenstate solutions of the spin-2 BEC Hamiltonian.  
Using the calibrated quantum lattice gas algorithm, the highly nonlinear physics in the nonintegrable regime of the spin-2 BEC is studied by performing soliton-soliton scattering experiments. The scattering of topological solitons produces breathers and complex quantum vortices characterized by  local entanglement across multiple $m_f$-hyperfine states of the Zeeman manifold. 
\end{abstract}

\keywords{spin-2 BEC, spinor superfluid, quantum entanglement, quantum simulation, quantum lattice gas, quantum computing, GPU parallel computing}

\maketitle

 \tableofcontents
\section{Introduction}
The  spinor superfluid phase of a spin-2 Bose-Einstein condensate (BEC) offers an opportunity to explore novel topological soliton structures that can emerge in quantum turbulence.  The spinor superfluid phase of a spin-2 BEC is of particular interest because of its ability to support complex quantum vortices, including non-Abelian quantum vortices \cite{kawaguchi:2010},  with  local entanglement across multiple $m_f$-states of the Zeeman manifold. The  reconnection process of complex quantum vortices is a fundamentally unitary process, and its affect on energy cascades remains an open area in quantum turbulence research.  
Analytical solutions of the spin-2 BEC Hamiltonian are reported here.  Also time-dependent quantum simulations of the spin-2 BEC in its spinor superfluid phase are presented here using a unitary quantum lattice gas algorithm found by one of our authors \cite{yepez:2016} that employs two spinor fermionic fields to model each spin-2 bosonic  field in the 5-dimensional Zeeman manifold---locally pairwise-entangled fermionic fields in the fermionic condensates represent the bosonic (order-parameter) field of the spin-2 BEC spinor superfluid.  One finding reported here is that the scattering of a pair of topological solitons produces breathers and complex quantum vortices which are characterized by  local entanglement across multiple $m_f$-hyperfine states of the Zeeman manifold.

The idea of a BEC was first proposed in 1924 by Bose \cite{Bose} to Einstein \cite{Einstein} and has markedly gained renewed interest following its first experimental realization 1995 at JILA \cite{Anderson}. This first man-made BEC was composed of Rubidium-87 atoms, which are spin-2 atoms, resulting in a condensate that is a spin-2 BEC.  However, the early experimentally realized 2-spinor BECs did not occupy the entire 5-dimensional hyperfine spin manifold (or Zeemann manifold) since the spin-2 BEC was trapped by a magnetic trap that forced the  weak-field seeking  Rubidium-87 atoms into, say, the $m_f=1,2$ levels of the manifold and excluding $m_f=0,-1,-2$.  More recently there have been experiments where the spin-2 BEC are entirely optically trapped for all the $m_f$ levels, which allows the Rubidium-87 atoms to simultaneously occupy the entire hyperfine manifold in quantum superposition \cite{chang2004}.  
To achieve the goal of experimentally verifying the production of a BEC superfluid, several experimental groups have demonstrated the ability to create quantum vortices in BEC's. Yet, today, the experimental search for specifically non-Abelian quantum vortices and the observation of their nonlinear interactions is still underway. 
 
 \subsection{Topological solition solutions}
 
Regarding analytical findings, we present exact analytic eigenstates for spin-2 BEC bright solitons and spin-2 BEC dark solitons in one dimension and Pad\'e approximant solutions in two spatial dimensions---solutions not  seen before by the authors.  In one-space dimension, BECs support topological  vortex solitons that come in two forms: (1) bright solitons with a high-density  region  in an otherwise near zero background (asymmptotically zero background) and (2) dark solitons with a low-density  region in an otherwise constant background. 
In two-space dimensions, such topological quantum vortex solitons can have a non-zero winding number, where phase of the condensate probability amplitude field accumulates in multiples of $2\pi$ as one traverses one full cycle around any closed contour that contains the vortex center.  Dark solitons were the first solitons to be produced experimentally and have now been produced by quite a few different groups \cite{Wieman, Matthews, Scherer, Haljan, Leslie, Burger, denschlag, becker} and bright solitons have been created in the laboratory by \cite{khaykovich, nguyen, marchant, strecker} among others.

 The bright and dark solitons of the 1D solutions presented here that quantum mechanically entangle three or more Zeeman levels have not been seen before by the authors.  Although, the two-level solutions of spin-2 BEC Hamiltonian were previously found \cite{zhang:2012}.   
  These dark soliton eigenstates in one-spatial dimension serve as the test state for the efficacy of the quantum lattice gas algorithm described in \cite{yepez:2016}---we can compare the numerically computed result with the analytically determined prediction.

Regarding the    dark soliton eigenstate solutions in two-space dimensions, we found these making use of the Pad\'e approximant method.  The Pad\'e approximant solution method was first used by Berloff  in scalar (spin-0) BEC superfluid \cite{Berloff, yepez_pade}.   Generalizaing this Pad\'e approximant solution method in a spin-2 BEC spinor superfluid is a principal analytical findings in this communication.  
With such analytical Pad\'e approximant  solutions available, comparisons of the  analytically and numerically predicted time-dependent solutions of the spin-2 BEC spinor superfluid equations two-spatial dimensions can be performed just as easily as in the one-dimensional case.  Such quantum simulation are carried out for the purpose of calibrating the quantum lattice gas algorithm and as a stepping stone to carrying out unitary quantum simulations of the soliton-soliton scattering experiments. 

\subsection{Soliton-soliton scattering via quantum simulation}

  The interaction of multiple quantum vortices---which have nonlinear reconnection physics---is analytically nonintegrable. Yet quantum simulation offers a way to faithfully capture the highly nonlinear dynamics of their rich mutual interactions \cite{yepez-vahala-qip-05,yepez-vahala-EPJ-09,PhysRevE.84.046713,yepez:084501}. 
A spin-2 BEC spinor superfluid supports complex quantum vortices (with local quantum entanglement with the spin-2 Zeeman manifold). 
  Complex solitons (including non-Abelian quantum vortices in two-dimensions) with local quantum entanglement in the Zeeman manifold can naturally emerge in during the time evolution  of spin-2 BEC spinor superfluid from simple initial conditions without any local entanglement. 
   These soliton-soliton scattering experiments produce complex solitons in both one- and two- spatial dimensions.  
 A number of quantum simulation examples are provided to demonstrate the common phenomena producing multiple complex solitons with local entanglement across multiple $m_f$-hyperfine states of the Zeeman manifold. 
   
Regarding the quantum simulation method from a computational mathematics perspective, the quantum lattice gas algorithm represents a dual Fermi condensate that is used to model the spin-2 BEC superfluid  \cite{yepez:2016}.  Moreover, the quantum lattice gas algorithm for the dual Fermi condensate---which has  a spin-2 BEC spinor superfluid phase---employs a novel operator splitting technique that mitigates the Baker-Campbell-Hausdorff catastrophe---basically splitting of the kinetic operator from the nonlinear and nondiagonal interaction operators into distinct dynamical subgroups.  Hence, the operator splitting the kinetic and nonlinear potential interaction occur without  error terms arising from noncommutivity up to fourth-order in an $\varepsilon$-expansion of the equation of motion.  Using the dual fermionic quantum lattice gas algorithm for numerical quantum simulation, the slope of the L2 norm error curve on a log-log plot is found to be $-4.77$, for example as measured using bright soliton solution states. 

 Yet, since there are two basic types of nonlinear-nondiagonal  operators that appear in the  interaction part of the spin-2 Hamiltonian\footnote{A derivation of these nonlinear and nondiagonal interaction operators is presented in the accompanying communication \cite{yepez:2016}.}, it is should be possible to improve the   numerical accurately of the quantum lattice gas algorithm by interleaving these particular interaction operators. Here we report that high numerical accurately of the quantum lattice gas algorithm can  indeed be obtained by interleaving these interaction operators: the slope of the L2 norm error curve on a log-log plot is favorably steeped to an observed value of $-5.67$ from the $-4.77$ value mentioned above.  The numerical convergence of the quantum lattice gas algorithm is robust in the bright soliton case.
 
   We also find  that for at least one class of soliton solutions---the kink (dark) soliton---that the unitary quantum lattice gas algorithm requires  interleaving  of its nonlinear-nondiagonal interaction operators to achieve high numerical convergence. For example, in the kink soliton case, we find  a slope of $-2.34$ for the  non-interleaved quantum algorithm and a much improve slope of  $-5.38$ for the interleaved quantum algorithm.    
 The overall verification of our operator-splitting technique represents a principle computational finding of our study of a spin-2 superfluid---we conclusively demonstrate that a strictly unitary quantum lattice gas algorithm can in fact accurately model the time-dependent evolution of a spin-2 BEC spinor superfluid in two-spatial dimensions including the generation and mutual interaction of complex quantum vortices.  

In summary, our numerical results  show that the quantum lattice gas algorithm agrees extremely well with the predicted analytical behavior and furthermore that it can converge to any chosen level of engineering precision.   
  Having established that the quantum lattice gas algorithm is  a faithful representation of the spin-2 BEC physics, we are able to study the mutual interactions of multiple complex solitons by the scattering two 
  analytical dark or bright soliton solutions in 1+1 dimensions and by scattering two analytical Pad\'e approximant dark soliton vortex solutions in 2+1 dimensions.

\subsection{Organization}

In Sec.~\ref{Section_The_theory_of_BEC_dynamics}, the theory of spinor BEC dynamics is presented.  We derive the equation of motion for spin-0, spin-1, and spin-2 BECs. These are the spin-2 Gross-Pitiaevskii equations which serve as the analytic comparison to our numerical simulations.    

In Sec.~\ref{Section_Solutions_to_the_spin_2_GPE_in_1D}, analytical solutions of the spin-2 Gross-Pitiaevskii equations  in 1+1 spacetime dimensions are presented.  This includes spatially flat, bright soliton solutions (hyperbolic secant), and dark soliton (hyperbolic tangent) energy eigenstate solutions to the spin-2 Gross-Pitiaevskii equations.  We find sixteen cases for each type of solution, seven of which we believe to be novel solutions.    

In Sec.~\ref{Section_Quantum_simulations_in_1D}, quantum simulations of a spin-2 BEC spinor superfluid  in 1+1 spacetime dimensions are presented.  We show that the analytic eigenstates of the spin-2 BEC behave like energy eigenstates when simulated using the quantum lattice gas algorithm.  We show that we can give the solitons momenta and have them move across the lattice. We show that the dispersion relation for the energies of moving solitons as calculated from the simulations are in agreement with analytically predicted energies.  Lastly we show collisions of solitons and observe that soliton collisions can create new solitons or breathers, excite unoccupied $m_f$ levels, and change which self interaction term is dominating the dynamics.  

In Sec.~\ref{Section_Solutions_to_the_Spinor_GPE_in_2D}, analytical Pad\'e approximant solutions of the spin-2 Gross-Pitiaevskii equations in 2+1 spacetime dimensions are presented.  We expand upon the Pad\'e approximant method used to solve the scalar BEC solution and apply it to a spin-2 BEC in two spatial dimensions.  We find eight different dark soliton solutions including three different solutions with multiple $m_f$ levels occupied.  We have not seen any of these solutions previously in the literature.  We also identify a condition on the Pad\'e approximant solutions to be more functionally similar to the hyperbolic tangent solutions of 1+1 spacetime dimensions.     

In Sec.~\ref{Section_Quantum_simulations_in_2D}, quantum simulations of a spin-2 BEC spinor superfluid  in 2+1 spacetime dimensions are presented.  We place four Pad\'e approximant eigenstate solutions in a quadrupole configuration to satisfy periodic boundary conditions.  We find that the Pad\'e approximant eigenstate solutions are indeed stationary when simulated using the quantum lattice gas algorithm.  These solutions can be given a momenta in any direction and we can collide solutions in different $m_f$ channels.  These collisions can excite dark soliton vortices in unoccupied channels which is a phenomena unique to spinor BEC.  We observe a conservation of winding number in each $m_f$ channel of the spin-2 BEC and track the paths of all the dark soliton vortices throughout the collision.

In Sec.~\ref{Section_Quantum_simulation_implementation}, some relevant implementation details regarding the quantum simulation method based on the quantum lattice gas algorithm of a spin-2 BEC spinor superfluid  is presented.  We show how to scale the algorithm to fit the numerical constraints of a simulation. We calculate the L2 norm for bright and dark soliton state to show the numerical convergence of the quantum lattice gas algorithm.  We introduce an operator interleaving procedure which can help improve the convergence of the dark soliton simulation. Lastly, we show the massive speed advantage  that implementing the quantum lattice gas algorithm on general-purpose graphics processing units provides.

In Sec.~\ref{Section_Conclusion}, a brief summary of the main findings are presented as well as some future outlooks for quantum simulations of a spin-2 BEC spinor superfluid, particular for the study of quantum turbulence.   
\section{The theory of spinor BEC dynamics}
\label{Section_The_theory_of_BEC_dynamics}

Presented is a derivation of the equations of motion for the spinor multiplet field (a set of coupled nonlinear partial differential equations for the $2f+1$ hyperfine states) of a spinor BEC.  The following derivation of the equation of motion for spin-2 BEC's largely follows the derivation from Kawaguchi and Ueda's excellent BEC review \cite{kawaguchi:2010}.
\subsection{Operator relations}
Before we derive the equation of motion we define and derive some operator relations to make our lives easier later on.  First, we define the field operators $\psi_m(r)$ and $\psi^\dagger_m(r)$ that are creation and annihilation operators for a bosonic particle with spin $f$  at position $r$ with quantum number $m$ (Zeeman $m_f$ level).  $\psi_m(r)$ and $\psi_m^\dagger(r)$ obey the usual equal-time commutation relations
\begin{subequations}
\label{commutationrelations}
\begin{align}
&\left[\psi_m(r), \psi_{m'}^\dagger(r')\right] = \delta_{mm'}\delta(r - r')\\
&\left[\psi_m(r), \psi_{m'}(r')\right] = \left[\psi_m^\dagger(r), \psi_{m'}^\dagger(r')\right] = 0.
\end{align}
\end{subequations}
We also define creation and annihilation operators for particle pairs $\hat{\bm{A}}_{F, M}\left(r, r'\right)$ and $\hat{\bm{A}}^\dagger_{F, M} \left( r, r'\right)$ that create or annihilate a pair of bosons at locations $r$ and $r'$ with a combined spin of $F$ and a combined spin in the $z$ direction of $M$. These can be related to the creation and annihilation operator of a single particle by the Clebsch-Gordon coefficients
\begin{widetext}
\begin{align}\label{CGcoeff}
\hat{\bm{A}}_{F, M}\left(r, r'\right) = \sum_{m, m' = -f}^{f}\braket{F, M | f, m; f, m'}\psi_m(r)\psi^\dagger_{m'}(r'). 
\end{align} 
\end{widetext}
Note that $\hat{\bm{A}}_{F, M}\left(r, r'\right) = 0$ if $F$ is odd. Since we are looking at spin-$f$ bosonic particles, $f$ will be integer spin thus a composite (of two particles each with spin $f$) must be an even integer. We will also define a total density operator $\hat{n}(r)$ as 
\begin{align}\label{noperator}
\hat{\bm{n}}(r) = \sum_{m = -f}^{f}\psi^\dagger_m(r)\psi_{m}(r), 
\end{align} 
a singlet pair operator as
\begin{align}\label{singlet}
\hat{\bm{A}}_{00}(r, r') = \frac{1}{\sqrt{2f + 1}} \sum_{m = -f}^{f}(-1)^{f-m}\psi_m(r)\psi_{-m}(r'),
\end{align}
and a spin density operator as
\begin{align}\label{spindensityoperator}
\hat{\bm{F}}_{\nu}(r) = \sum_{m, m' = -f}^{f} (\bm{f}_\nu)_{mm'}\psi^\dagger_m(r)\psi_{m'}(r)
,
\end{align}
where $\nu$ is a cartesian coordinate either $x$, $y$, or $z$.  Lastly, we need the projection operator, $\hat{\mathcal{P}}_F$, onto a two-body state with total spin $F$
\begin{align}
\hat{\mathcal{P}}_F =  \sum_{M = -F}^{F}|{F, M}\rangle \langle {M, F} |\label{projection}, 
\end{align}
where
\begin{align}
\sum_{F}\hat{\mathcal{P}}_F = I, \label{completeness}
\end{align}
and where $I$ is an $F$-dimensional identity matrix. Acting $\psi^\dagger_{m_1}(r)\psi^\dagger_{m_2}(r')\langle{f, m_1; f, m_2}|$ on the left and $|{f, m'_1; f, m'_2}\rangle\psi_{m'_1}(r')\psi_{m'_2}(r)$ on the right of  (\ref{completeness}), we get 
\begin{align}\label{nterm}
: \hat{\bm{n}}(r) \hat{\bm{n}}(r') : = \sum_{F = 0}^{F = 2f} \sum_{M = -F}^{M=F}  \hat{\bm{A}}^\dagger_{F, M}\left(r, r'\right) \hat{\bm{A}}_{F, M}\left(r, r'\right),
\end{align}
where  :: denotes normal ordering and the summation over $F$ is only over even numbers since $ \hat{\bm{A}}_{F, M}\left(r, r'\right) = 0$ for odd $F$.  The last relation we need comes from the composition law for angular momentum
\begin{align}\label{complaw}
\bm{f}_1 \cdot \bm {f}_2 = \frac{1}{2}\left(\left(\bm{f}_1 + \bm {f}_2 \right)^2 -\bm{f}^2_1 - \bm{f}^2_2\right) = \frac{1}{2}\bm{f}^2_{total} - f(f+1),
\end{align}
where the $\bm{f}_{1,2}$ are angular momentum operators acting on particles of total spin $f$ and $\bm{f}_{total} = \bm{f}_1 + \bm {f}_2$.  Now acting $\psi^\dagger_{m_1}(r)\psi^\dagger_{m_2}(r')\langle{f, m_1; f, m_2}|$ on the left and $|{f, m'_1; f, m'_2}\rangle\psi_{m'_1}(r')\psi_{m'_2}(r)$ on the right of  (\ref{complaw}) we get 
\begin{multline}\label{generalspindensity}
:\hat{\bm{F}}(r) \cdot \hat{\bm{F}}(r'): = \sum_{F=0}^{F = 2f}\left(\frac{1}{2}F\left(F+1\right) - f\left(f+1\right)\right)\\
\times\sum^F_{M= -F}\hat{\bm{A}}^\dagger_{F,M}\left(r, r'\right) \hat{\bm{A}}_{F, M}\left(r, r'\right).
\end{multline}
\subsection{Equation of motion}
The Hamiltonian for a spin-$F$ BEC can be split into two parts, a noninteracting diagonal part $H_{diag}$ and an interacting part $H_{int}$ such that the total hamiltonian  $H$ is given by 
\begin{align}\label{hamilton}
H = H_{diag} + H_{int}.
\end{align}
In this derivation we will restrict our focus to BEC's in the absence of an external potential, thus 
\begin{align}\label{hamiltonkinetic}
H_{diag} = \int{dr\sum_{m_f}\psi_m^\dagger(r)\left(-\nabla^2\right)\psi_m(r)},
\end{align}
where we have set $\hbar = 1$ and $m = 1/2$. The interaction Hamiltonian for 2 particles with total spin $F$ is given by 
\begin{align}\label{interactionHam}
H_{int}^{F} = \int{dr\int{dr' \frac{1}{2}v_{F}(r, r') \sum_{M = -F}^{M = F} \hat{\bm{A}}_{F, M}^\dagger(r, r') \hat{\bm{A}}_{F, M}(r, r')}} 
\end{align}
where $v_{F}(r, r')$ is the energy between the two bosons at $r$ and $r'$ which we will approximate by an effective coupling constant $\gamma_F$ multiplying a delta function
\begin{align}
v_F(r, r') = \gamma_F\delta(r - r').
\end{align} 
Once we have the hamiltonian we minimize the energy functional in (\ref{hamiltonkinetic}) and (\ref{interactionHam}) to get the equation of motion 
\begin{align}\label{eomderiv}
i\hbar\frac{\partial \psi_m(r)}{\partial t} = \frac{\partial E}{\partial \psi^*_m(r)} = \frac{\partial \langle H \rangle}{\partial \psi^*_m(r)}
.
\end{align}
\subsubsection{Spin 0}
For the spin-0 case, after the $dr'$ integration, the interaction Hamiltonian is  given by
\begin{align} 
H_{int}^0 = \frac{\gamma_0}{2}\int{dr \hat{\bm{A}}_{0, 0}^\dagger(r, r) \hat{\bm{A}}_{0, 0}(r, r)},
\end{align} 
where the spin-singlet terms are
\begin{align}
\hat{\bm{A}}_{0, 0}(r, r) = \psi_0(r)\psi_0(r) 
\end{align}
and
\begin{align}
\hat{\bm{A}}^\dagger_{0, 0}(r, r) = \psi_0^*(r)\psi_0^*(r).
\end{align}
Thus, the energy in the mean field is given by 
\begin{align}\label{hamspin0}
E = \langle H \rangle = \int{\psi_0^*(r)\left(-\nabla^2\right)\psi_0(r)  + \frac{\gamma_0}{2}\psi_0^*(r)^2 \psi_0(r)^2 dr},
\end{align}
and the equation of motion is given by substituting  (\ref{hamspin0}) into  (\ref{eomderiv}), which gives
\begin{align}
i\hbar\frac{\partial \psi_0(r)}{\partial t} = \left(-\nabla^2 + \gamma_0\left |\psi_0(r) \right |^2 \right)\psi_0(r)
.
\end{align}
\subsubsection{Spin 1}
For the spin-1 case, the interaction Hamiltonian is given by 
\begin{align}
H_{int} = H^0_{int} + H^2_{int},
\end{align}
where
\begin{align} 
H^0_{int} = \frac{\gamma_0}{2}\int{dr \hat{\bm{A}}_{0, 0}^\dagger(r, r) \hat{\bm{A}}_{0, 0}(r, r)}
\end{align} 
and 
\begin{align} 
H^2_{int} = \frac{\gamma_2}{2}\int{dr \sum_{-M}^{M}\hat{\bm{A}}_{2, M}^\dagger(r, r) \hat{\bm{A}}_{2, M}(r, r)}
.
\end{align}
Upon comparison to the $f=1$ versions of  (\ref{nterm}) and  (\ref{generalspindensity}) we see that the full interaction Hamiltonian for a spin-1 BEC is  
\begin{align}\label{hamspin1}
H_{int} = \frac{1}{2} \int{g_0 : \hat{\bm{n}}(r) \hat{\bm{n}}(r) :  +  g_1 :\hat{\bm{F}}(r) \cdot \hat{\bm{F}}(r'):},
\end{align}
where 
\begin{align}
g_0 =  \frac{\gamma_{0} + 2 \gamma_{2}}{3}
\qquad
g_1 =  \frac{\gamma_{2} - 2 \gamma_{0}}{3}.
\end{align}
The equation of motion is given by substituting  (\ref{hamspin1}) into  (\ref{eomderiv}), which gives
\begin{align}
\label{eomspin1}
i\hbar\frac{\partial \psi_m(r)}{\partial t} = \left(-\nabla^2 + g_0 \left |\psi(r) \right |^2 \right)\psi_m(r) 
+g_1
\bm{\hat{F}} \cdot \bm{f}\psi_m(r),
\end{align}
where $\bm{f} = (f_x, f_y, f_z)$, and $f_x, f_y, $ and $f_z$ are given by their usual spin-1 $SU(2)$ representations
\begin{subequations}
\begin{align}
f_x =
\frac{\hbar}{\sqrt{2}}
{\scriptsize
\begin{pmatrix}
  0    & 1 & 0  \\
  1    & 0 & 1 \\
  0  & 1 & 0 \\
\end{pmatrix}
},
\;\;
f_y =
 \frac{i \hbar}{\sqrt{2}}
{\scriptsize
\begin{pmatrix}
  0    & -1 & 0 \\
  1   & 0 & -1 \\
  0  & 1 &  0 \\
\end{pmatrix}
},
\;\;
f_z =
\hbar
{\scriptsize
\begin{pmatrix}
  1   & 0 & 0  \\
  0   & 0 & 0  \\
  0  & 0 & -1 \\
\end{pmatrix}
}.
\end{align}
\end{subequations}
\subsubsection{Spin 2}
For the spin-2 case, the interaction Hamiltonian is given by 
\begin{align}
H_{int} = H^0_{int} + H^2_{int} + H^4_{int}
,
\end{align}
where
\begin{align} 
&H^0_{int} = \frac{\gamma_0}{2}\int{dr \hat{\bm{A}}_{0, 0}^\dagger(r, r) \hat{\bm{A}}_{0, 0}(r, r)}\\
&H^2_{int} = \frac{\gamma_2}{2}\int{dr \sum_{M = -2}^{M = 2}\hat{\bm{A}}_{2, M}^\dagger(r, r) \hat{\bm{A}}_{2, M}(r, r)}
\end{align}
and
\begin{align} 
H^4_{int} = \frac{\gamma_4}{2}\int{dr \sum_{M = -4}^{M = 4}\hat{\bm{A}}_{4, M}^\dagger(r, r) \hat{\bm{A}}_{4, M}(r, r)}
.
\end{align}
Upon comparison to the $f=2$ versions of  (\ref{nterm}) and  (\ref{generalspindensity}) we see that the full interaction Hamiltonian for a spin-2 BEC is  
\begin{align}
\nonumber
H_{int} = \frac{1}{2} \int{g_0 : \hat{\bm{n}}(r) \hat{\bm{n}}(r) :  +  g_1 :\hat{\bm{F}}(r) \cdot \hat{\bm{F}}(r'):}  \\ 
+ g_2 \hat{\bm{A}}_{0, 0}^\dagger(r, r) \hat{\bm{A}}_{0, 0}(r, r),
\label{hamspin2}
\end{align}
where 
\begin{align}
g_0 =  \frac{4\gamma_{2} + 3 \gamma_{4}}{7},
\qquad
g_1 =  \frac{\gamma_{4} -  \gamma_{2}}{7},
\qquad
g_2 =  \frac{7\gamma_{0} - 10 \gamma_{2} + 3 \gamma_{4}}{7}.
\end{align}
The equation of motion is given by substituting  (\ref{hamspin2}) into  (\ref{eomderiv}), which gives
\begin{multline}\label{eomspin2}
i\frac{\partial \psi_m(r)}{\partial t} = \left(-\nabla^2 + g_0 \left |\psi(r) \right |^2 \right)\psi_m(r) 
\\+g_1\bm{\hat{F}} \cdot \bm{f}\psi_m(r)+ g_2 \left|{A}_{00}\right|^2\psi^*_{-m}(r), 
\end{multline}
where
\begin{align}\label{spinsinglet}
\left|{A}_{00}\right|^2 = \frac{1}{5}\left(2\psi_(x)2\psi_{-2}(x) - 2\psi_1(x)\psi_{-1}(x) + \psi_0^2(x)\right), 
\end{align}
and where now $f_x, f_y, $ and $f_z$ are given by their usual spin-2 $SU(2)$ representations
\begin{widetext}
\begin{align}
& f_x =
\hbar
{\scriptsize
\begin{pmatrix}
  0    & 1 & 0 & 0 &  0  \\
  1    & 0 & \sqrt{\frac{3}{2}} & 0 & 0 \\
  0  &  \sqrt{\frac{3}{2}}  & 0 &  \sqrt{\frac{3}{2}}  & 0 \\
  0 & 0 &  \sqrt{\frac{3}{2}}  & 0 & 1 \\
  0 & 0 & 0 & 1 & 0\\
\end{pmatrix},
}
\qquad
f_y =
i
\hbar
{\scriptsize
\begin{pmatrix}
  0    & -1 & 0 & 0 & 0   \\
  1   & 0 & -\sqrt{\frac{3}{2}} & 0 & 0 \\
  0  & \sqrt{\frac{3}{2}}  & 0 & -\sqrt{\frac{3}{2}}  &  0 \\
  0 & 0 & \sqrt{\frac{3}{2}}  & 0 & -1\\
  0& 0&0 & 1 & 0\\ 
\end{pmatrix},
} 
\qquad
f_z =
\hbar
\begin{pmatrix}
  2   & 0 & 0 & 0 & 0   \\
  0   & 1 & 0 & 0 & 0 \\
  0  & 0 & 0 & 0 & 0 \\
  0 & 0 & 0 & -1 & 0 \\
  0 & 0 & 0 & 0 &-2
\end{pmatrix}
.
\end{align}
The full equation of motion  for a spin-2 BEC in its component form is given by
\begin{align}\label{eom_component_form}
i \hbar \partial_t
\begin{pmatrix}
\psi_{2}\\
\psi_{1}\\
\psi_{0}\\
\psi_{-1}\\
\psi_{-2}\\
\end{pmatrix}
=
 \begin{pmatrix}
(-\frac{\hbar^2\nabla^2}{2m}  +  g_0 {\rho})\psi_{2} + g_1 (\bm{\hat{F}} \cdot \bm{f})_{2}  +g_2|A_{00}|^2\psi^*_{-2}  \\
(-\frac{\hbar^2\nabla^2}{2m}  + g_0 {\rho})\psi_{1} + g_1 (\bm{\hat{F}} \cdot \bm{f})_{1}+g_2|A_{00}|^2\psi^*_{-1}  \\
(-\frac{\hbar^2\nabla^2}{2m}  + g_0 {\rho})\psi_{0} + g_1 (\bm{\hat{F}} \cdot \bm{f})_{0}  +g_2|A_{00}|^2\psi^*_{0}  \\
(-\frac{\hbar^2\nabla^2}{2m}  + g_0 {\rho})\psi_{-1} + g_1 (\bm{\hat{F}} \cdot \bm{f})_{-1} +g_2|A_{00}|^2\psi^*_{1}  \\
(-\frac{\hbar^2\nabla^2}{2m}  + g_0 {\rho})\psi_{-2} + g_1(\bm{\hat{F}} \cdot \bm{f})_{-2}  +g_2|A_{00}|^2\psi^*_{2}  \\
\end{pmatrix},
\end{align}
where
\begin{subequations}
\label{definitions}
\begin{align}
 \rho = \sum_{m}{|\psi_m|^2}, 
 \end{align}
and
\begin{align}
(\bm{\hat{F}} \cdot \bm{f})_{\pm 2} 
&= 4 \psi_{\pm 2}^3 - 4 \psi_{\mp 2}^2 \psi_{\pm 2}-2 \psi_{\mp 1}^2 \psi_{\pm 2}+4 \psi_{\pm 1}^2 \psi_{\pm 2}
 + 2 \psi_{\mp 2} \psi_{\mp 1} \psi_{\pm 1}+\sqrt{6} \psi_0 \psi_{\pm 1} \left(c_{\mp 1}+\psi_{\pm 1}\right), 
 \\ 
 (\bm{\hat{F}} \cdot \bm{f})_{\pm 1} 
 &=  \psi_{\pm 1}^3 +\left(-2 \psi_{\mp 2}^2-\psi_{\mp 1}^2+3 \psi_0^2+4 \psi_{\pm 2}^2+2 \sqrt{6} \psi_0 \psi_{\pm 2}\right) \psi_{\pm 1} + \psi_{\mp 1} \left(\psi_{\mp 2} \left(\sqrt{6} \psi_0+2 \psi_{\pm 2}\right)+\psi_0 \left(3 \psi_0+\sqrt{6} \psi_{\pm 2}\right)\right),  
\\ 
 (\bm{\hat{F}} \cdot \bm{f})_{0}
 &= \left(\psi_{-1}+\psi_1\right) \left(3 \psi_0 \left(\psi_{-1}+\psi_1\right)+\sqrt{6} \left(\psi_{-2} \psi_{-1}+\psi_1 \psi_2\right)\right).
\end{align}
\end{subequations}
\end{widetext}
From this point on we use natural units and set $\hbar = 1$ and $m = 1/2$. 
\subsection{Additional terms}
In addition to the kinetic and self-interaction terms that appear in (\ref{eom_component_form}) the equation of motion for spin-2 BECs can also include terms coming from an external potential, often times a magnetic field, or a chemical potential term of the form $\mu \cdot \psi_m$.  We will limit ourselves to the case where there is no external potential, and will only briefly mention the chemical potential during our discussion of soliton energies.

\section{Solutions to the spin-2 GP equation in 1D}
\label{Section_Solutions_to_the_spin_2_GPE_in_1D}

\subsection{Thomas-Fermi solutions}
The first set of solutions we consider are the solutions that arise from the Thomas-Fermi approximation.  This approximation assumes a uniform flat quantum fluid such that the spatial derivative in  (\ref{eomspin2}) is identically zero everywhere.  In other words we assume the quantum field has the form $\psi_{m}(x, t) = c_m e^{-i E t}$.  Substituting this in for into  (\ref{eomspin2}) gives the following system of five equations,
\begin{align}
\label{system_of_equations}
\begin{pmatrix}
0\\
0\\
0\\
0\\
0\\
\end{pmatrix}
= \begin{pmatrix}
-c_2 E + {\rho} c_2g_0 + \xi_2 g_1 +|A_{00}|^2 c_{-2}g_2\\
-c_1 E + {\rho}c_1g_0 + \xi_1 g_1 +|A_{00}|^2 c_{-1}g_2\\
-c_0 E + {\rho} c_0g_0 + \xi_0 g_1 + |A_{00}|^2 c_{0}g_2\\
- c_{-1} E + {\rho} c_{-1}g_0 + \xi_{-1} g_1 +|A_{00}|^2 c_{1}g_2\\
- c_{-2} E + {\rho} c_{-2}g_0 + \xi_{-2} g_1 +|A_{00}|^2 c_{2}g_2\\
\end{pmatrix},
\end{align}
where
\begin{align}\label{alphabeta}
\nonumber
\xi_{\pm 2} &= 4 c_{\pm 2}^3-4 c_{\mp 2}^2 c_{\pm 2}-2 c_{\mp 1}^2 c_{\pm 2}+4 c_{\pm 1}^2 c_{\pm 2}, \\ 
&+ 2 c_{\mp 2} c_{\mp 1} c_{\pm 1}+\sqrt{6} c_0 c_{\pm 1} \left(c_{\mp 1}+c_{\pm 1}\right), 
\end{align} 
\begin{align}
\nonumber
 \xi_{\pm 1} &=  c_{\pm 1}^3+\left(-2 c_{\mp 2}^2-c_{\mp 1}^2+3 c_0^2+4 c_{\pm 2}^2+2 \sqrt{6} c_0 c_{\pm 2}\right) c_{\pm 1}, \\ 
&+c_{\mp 1} \left(c_{\mp 2} \left(\sqrt{6} c_0+2 c_{\pm 2}\right)+c_0 \left(3 c_0+\sqrt{6} c_{\pm 2}\right)\right)  
\end{align} 
\begin{align}
 \xi_{0} &= \left(c_{-1}+c_1\right) \left(3 c_0 \left(c_{-1}+c_1\right)+\sqrt{6} \left(c_{-2} c_{-1}+c_1 c_2\right)\right),
\end{align}
and  $\rho$ and $|A_{00}|^2$ retain their definitions from (\ref{definitions}) and (\ref{spinsinglet}) respectively.  This is still a rather cumbersome set of equations, and the general solution is not yet known since the equation is nonlinear in the $c_m$ variables.  However, we have found sixteen different non-trivial solutions, three of which have non-zero fields in all five components. These solutions are enumerated in Table~\ref{Table_I_1D_solutions}.  
\paragraph*{}
The trick to finding these solutions is to make assumptions about the $c_m$ coefficients that simplify the system of equations.  For the solutions nos. $1 - 9$ in Table~\ref{Table_I_1D_solutions}, any $c_m$ that is equal to zero was assumed to be zero before solving (\ref{system_of_equations}).  For solutions nos. $10 - 16$, the quantum field $\psi_m(x)$ is assumed to have the respective form 
\begin{align}
&\hspace{1cm} \psi_{10} 
=
{\footnotesize
 \begin{pmatrix}
c_2\\
0\\
c_0\\
0\\
-c_2\\
\end{pmatrix}
},
\text{  }
\psi_{11} 
=
{\footnotesize
 \begin{pmatrix}
c_2\\
0\\
c_0\\
0\\
c_2\\
\end{pmatrix}
},
\text{  }
\psi_{12} 
= 
{\footnotesize
\begin{pmatrix}
0\\
c_1\\
c_0\\
-c_1\\
0\\
\end{pmatrix}
},
\nonumber
\\ 
&\psi_{13} 
= 
{\footnotesize
\begin{pmatrix}
c_2\\
c_1\\
0\\
-c_1\\
-c_2\\
\end{pmatrix}
},
\text{  }
\psi_{14} 
= 
{\footnotesize
\begin{pmatrix}
c_2\\
c_1\\
c_0\\
c_1\\
c_2\\
\end{pmatrix}
},
\text{  }
\psi 
= 
 {\footnotesize
\begin{pmatrix}
c_2\\
c_1\\
c_0\\
c_1\\
c_2\\
\end{pmatrix}
},
\text{  }
\psi_{15} 
= 
{\footnotesize
\begin{pmatrix}
c_2\\
c_1\\
c_0\\
-c_1\\
c_2\\
\end{pmatrix}
}.
\label{entanglement_structure}
\end{align}

As a warm-up example, let us first solve for solution no. 1 in Table~\ref{Table_I_1D_solutions}. We start with 
\begin{align}
\psi_{16} 
=
{\footnotesize
 \begin{pmatrix}
c_2\\
0\\
0\\
0\\
0\\
\end{pmatrix}},
\end{align}
which when substituted into  (\ref{system_of_equations}) gives 
\begin{align}
&0 = -c_2 E + \rho c_2 g_0 + 4 c_2^3 g_1,  
\end{align}
and after substituting in for ${\rho}$ yields
\begin{align}
&0 = -c_2 E  + c_2^3 g_0 + 4 c_2^3 g_1 
\implies 0 = -E + c_2^2(g_0 + 4 g_1) 
,
\end{align}
which gives the solution 
\begin{align}
c_2 = \pm \sqrt{\frac{E}{g_0 + 4g_1}}.
\end{align}
As a more challenging example, let us solve for solution no. $16$.  We begin by assuming $\psi$ is of the form 
\begin{align} 
\psi 
= 
{\footnotesize
\begin{pmatrix}
c_2\\
c_1\\
c_0\\
- c_1\\
c_2\\
\end{pmatrix}}.
\end{align}
This simplifies  (\ref{system_of_equations}) to 
\begin{align}{
\label{tf_soe}
\begin{pmatrix}
0\\
0\\
0\\
0\\
0\\
\end{pmatrix}
= \begin{pmatrix}
-c_{2} E + \rho c_{2}g_0 +|A_{00}| c_{2}g_2\\
-c_1 E + \rho c_1g_0  -|A_{00}| c_{1}g_2\\
-c_0 E + \rho c_0g_0 +|A_{00}|c_{0}g_2\\
c_1 E - \rho c_1g_0  +|A_{00}|c_{1}g_2\\
-c_{2} E + \rho c_{2}g_0 +|A_{00}|c_{2}g_2\\
\end{pmatrix}}.
\end{align}
A closer examination of  (\ref{tf_soe}) reveals that there is only one independent equation, which is
\begin{align}
\label{one_eq}
E =  \rho g_0 + |A_{00}| g_2 .
\end{align} 
Since we still have three independent variables $c_2$, $c_1$, and $c_0$, the best we can do is find a family of solutions where we have one coefficient in terms of the other two.  We will choose to express $c_0$ in terms of $c_1$ and $c_2$.  So substituting in $\alpha$ and $\beta$ into  (\ref{one_eq}) we get 
\begin{align}
&E = \left(2c_{2}^2+2c_{1}^2+c_0^2\right)g_0  + \frac{1}{5}\left(2c_{2}^2 + 2c_{1}^2+c_0^2 \right)g_2 
\end{align} 
\begin{align}
&E - 2 \left (c_2^2 + c_1^2 \right)\left(g_0 + \frac{g_2}{5} \right)  = c_0^2\left(g_0 + \frac{g_2}{5} \right) 
\end{align} 
\begin{align}
&5 E - 2 \left (c_2^2 + c_1^2 \right)\left(5g_0 + g_2 \right)  = c_0^2\left(5 g_0 + g_2 \right) 
\end{align} 
\begin{align}
&\frac{5 E - 2 \left (c_2^2 + c_1^2 \right)\left(5g_0 + g_2 \right)}{\left(5 g_0 + g_2 \right)}  = c_0^2  
\end{align} 
\begin{align}
& c_0 = \pm\sqrt{\frac{5 E -  \left (c_2^2 + c_1^2 \right)\left(10g_0 + 2g_2 \right)}{\left(5 g_0 + g_2 \right)} } ,
\end{align} 
which is solution no. 16 in Table~\ref{Table_I_1D_solutions}.   All the other solutions in the table can be solved for in similar fashion, and are presented in Table~\ref{Table_I_1D_solutions}.  It is important to note that since we have assumed that all the $c_m$ are real the Thomas-Fermi solutions are only valid in the regime where the $g_i$ produce real valued $c_m$. 
\begin{table}[!h!t!b!p]
\scriptsize
\begin{center}
\begin{tabular}{ |c|c|c|c|c|c|}
\hline
\multicolumn{6}{|c|}{1D Solutions} \\
\hline
 & $c_2$ & $c_1$ & $c_0$ & $c_{-1}$ &$c_{-2}$\\
\hline 
0 & 0 & 0 & 0 & 0 & 0 \\[.2cm]
1 &  $\sqrt{\frac{E}{g_0+4 g_1}}$ & 0 & 0 & 0 & 0 \\[.2cm]
2 &  0 & 0 & 0 & 0 & $\sqrt{\frac{E}{g_0+4 g_1}}$ \\[.2cm]
3 & 0 & $\sqrt{\frac{E}{g_0+g_1}}$ & 0 & 0 & 0 \\[.2cm]
4 & 0 & 0 & 0 & $\sqrt{\frac{E}{g_0+g_1}}$ & 0 \\[.2cm]
5 & 0 &  0 & $\sqrt{\frac{5 E}{5g_0+g_2}}$  & 0 & 0 \\[.2cm]
\hline & & & & &\\ 
6 &  $\sqrt{\frac{5 E}{10g_0+2g_2}}$ & 0 & 0 & 0 &  $\sqrt{\frac{5 E}{10g_0+2g_2}}$ \\[.2cm]
7 & 0 & $\sqrt{\frac{5 E}{10g_0+2g_2}}$ & 0 & $\sqrt{\frac{5 E}{10g_0+2g_2}}$ & 0 \\[.2cm]
8  &  $\sqrt{\frac{E}{3g_0}}$ & 0 & 0 &  $\sqrt{\frac{2 E}{3g_0}}$ & 0 \\[.2cm]
9  &  0 & $\sqrt{\frac{2 E}{3g_0}}$ & 0 &  0 & $\sqrt{\frac{E}{3g_0}}$  \\[.2cm]
\hline & & & & &\\ 
10 & $\sqrt{\frac{E} {4g_0}}$ & 0 & $\sqrt{\frac{E} {2g_0}}$ & 0 & $ -\sqrt{\frac{E} {4g_0}}$ \\[.2cm]
11 & $c_{2}$ & 0 & $\sqrt{\frac{5 E - 10 c_{2}^2 g_0 - 2 c_{2}^2 g_2}{5 g_0+g_2}}$ & 0 & ${c_2}$ \\[.2cm]
12 & 0 & $c_1$ & $\sqrt{\frac{5 E - 10 c_{1}^2 g_0 - 2 c_{1}^2 g_2}{5 g_0+g_2}}$ & $-c_{1}$ & 0 \\[.2cm]
13 &  $\sqrt{\frac{E}{4g_0+4g_1}}$ &  $\sqrt{\frac{E}{4g_0+4g_1}}$ & 0 &  $-\sqrt{\frac{E}{4g_0+4g_1}}$ &  $-\sqrt{\frac{E}{4g_0+4g_1}}$ \\[.2cm]
14 & $\sqrt{\frac{3 E}{16 g_0}}$ & $\sqrt{\frac{E}{4 g_0}}$& $-\sqrt{\frac{E}{ 8 g_0}}$ & $\sqrt{\frac{E}{4 g_0}}$ & $\sqrt{\frac{3 E}{16 g_0}}$ \\[.2cm]
15 & $\frac{1}{4} \sqrt{\frac{E}{g_0+4 g_1}}$ & $\frac{1}{2} \sqrt{\frac{E}{g_0+4 g_1}}$ & $\frac{1}{2}\sqrt{\frac{3 E}{2 g_0+8 g_1}}$ & $\frac{1}{2} \sqrt{\frac{E}{g_0+4 g_1}}$ & $\frac{1}{4} \sqrt{\frac{E}{g_0+4 g_1}}$ \\[.2cm]
16 & $c_2$ & $c_1$ & $\sqrt{\frac{5 E - \left(c_2^2+c_1^2\right) \left(10 g_0+2 g_2\right)}{5 g_0+g_2}}$  & $-c_1$ & $c_2$ \\[.2cm]
\hline
\end{tabular}
\end{center}
\caption{\label{Table_I_1D_solutions} \footnotesize  The coefficients that solve the spin-2 BEC equation of motion given the Thomas-Fermi approximation.  In the process of deriving these solutions we assume that the $c_m = c_m^*$ so the Thomas-Fermi solutions only apply where the values of $g_0$, $g_1$, and $g_2$ are such that all the $c_m$ are real.  Solution nos. 6 through 16 all represent the asymptotic states of topological solitons with local pairwise entanglement within the 5-dimensional Zeeman manifold of a spin-2 superfluid. For solutions nos. 10-16 this entanglement was imposed a priori by the ansatz made in (\ref{entanglement_structure}).These are also the coefficients for bight soliton, $\psi_m (x) = c_m\Sech(kx)$, and the dark soliton, $\psi_m (x) = c_m\Tanh(kx)$ solutions. For bright solitons solutions $E = -2$ and for dark solitons $E = 2$.}
\end{table}

\subsection{Local quantum entanglement}

This is a convenient opportunity to remind ourselves that the quantum lattice gas algorithm is capable of being implemented on a quantum computer.  Thus we could initialize an entangled qbit array to ensure that the delicate balance between the $m_f$ levels is maintained.  The spin-2 quantum lattice gas algorithm requires 10 qbits per lattice site as described in \cite{yepez:2016}.  In general each lattice site can be represented by
\begin{align}
\nonumber
|{\psi}\rangle =& \sum_{q_1 = 0}^{1} \,
\sum_{q_2 = 0}^{1} \cdots 
\\
\label{full_state}
&
\sum_{q_{10} = 0}^{1} 
\mathcal{A}(q_1, q_2, \dots , q_{10})
{ | 
\underbracket{q_1q_2}_{2}\,
\underbracket{q_3q_4}_{1}\,
\underbracket{q_5q_6}_{0}\,
\underbracket{q_7q_8}_{-1}\,
\underbracket{q_9q_{10}}_{-2} \rangle},
\end{align}
where, the underbraces connect the 2 qbits representing a single $m_f$ level of the spinor BEC and $\mathcal{A}(q_1, q_2, \dots , q_{10}) $ are normalization coefficients such that  
\begin{align}
 \sum_{q_1 = 0}^{1} \,
\sum_{q_2 = 0}^{1} \cdots \sum_{q_{10} = 0}^{1} 
 \mathcal{A}(q_1, q_2, \dots , q_{10})  = 1.    
\end{align} 
The entangled versions of the solutions in Table~\ref{Table_I_1D_solutions} can be written as 
\begin{widetext}
\begin{subequations}
\label{entangled_states}
\begin{align}
&\ket{\psi_6} = \sqrt{1 - \alpha^2}\ket{0000000000} + \alpha\ket{1100000011} 
,\qquad \alpha(x) = f(x)\text{\footnotesize$\sqrt{\frac{5 E}{10g_0+2g_2}}$}
\\ 
&\ket{\psi_7} = \sqrt{1 - \alpha^2}\ket{0000000000} + \alpha\ket{0011001100} 
,\qquad \alpha(x) = f(x)\text{\footnotesize$\sqrt{\frac{5 E}{10g_0+2g_2}}$}
\\ 
&\ket{\psi_8} = \sqrt{1 - \alpha^2}\ket{0000000000} + \alpha(\ket{1100000000} + \sqrt{2} \ket{0000001100}) 
,\qquad \alpha(x) = f(x)\text{\footnotesize$\sqrt{\frac{E}{3g_0}}$}
\\
&\ket{\psi_9} = \sqrt{1 - \alpha^2}\ket{0000000000} + \alpha(\ket{0011000000} + \sqrt{2} \ket{0000000011}) 
,\qquad \alpha(x) = f(x)\text{\footnotesize$\sqrt{\frac{E}{3g_0}}$}
\end{align}
\begin{align}
&\ket{\psi_{10}} = \sqrt{1 - \alpha^2}\ket{0000000000} + \alpha(\ket{1100000000} + \sqrt{2} \ket{0000110000} - \ket{0000000011})
,\qquad \alpha(x) = f(x)\text{\footnotesize$\sqrt{\frac{E}{4g_0}}$}
\end{align}
\begin{align}
&\ket{\psi_{11}} = \sqrt{1 - \alpha^2 - \beta^2}\ket{0000000000} + \alpha \ket{1100000011} + \beta \ket{0000110000})
,\qquad \beta(x) = f(x)\text{\footnotesize$\sqrt{\frac{5 E}{5g_0 + g_2} - 2 \alpha^2(x)}$}
\end{align}
\begin{align}
&\ket{\psi_{12}} = \sqrt{1 - 2 \alpha^2 - \beta^2}\ket{0000000000} + \alpha \ket{0011000000} + \beta \ket{0000110000} - \alpha \ket{0000001100} )
,\quad \beta(x) = f(x)\text{\footnotesize$\sqrt{\frac{5 E}{5g_0 + g_2} - 2 \alpha^2(x)}$}
\end{align}
\begin{align}
&\ket{\psi_{13}} = \sqrt{1 - \alpha^2}\ket{0000000000} + \alpha(\ket{1111000000} - \ket{0000001111}) 
,\qquad \alpha(x) = f(x)\text{\footnotesize$\sqrt{\frac{E}{4 g_0+ 4 g_1}}$}
\\ 
&\ket{\psi_{14}} = \sqrt{1 - \alpha^2}\ket{0000000000} + \alpha(\text{\footnotesize{$\sqrt{\frac{3}{2}}$}}\ket{1100000011} - \ket{0000110000} + \sqrt{2}\ket{0011001100})
,\qquad \alpha(x) = f(x)\text{\footnotesize$\sqrt{\frac{E}{8 g_0}}$}
\\ 
&\ket{\psi_{15}} = \sqrt{1 - \alpha^2}\ket{0000000000} + \alpha(\ket{1100000011} + \sqrt{6}\ket{0000110000} +
2 \ket{0011001100})
,\qquad \alpha(x) = f(x)\text{\footnotesize$\frac{1}{4}\sqrt{\frac{5 E}{10g_0+2g_2}}$}
\\ 
&\ket{\psi_{16}} = \sqrt{1 - \alpha^2 - 2 \beta^2 - \gamma^2}\ket{0000000000} + \alpha \ket{1100000011} + \beta \ket{0011000000} + \gamma \ket{0000110000} - \beta \ket{0000001100}),\\ \nonumber
& \qquad \gamma(x) = f(x)\text{\footnotesize$\sqrt{\frac{5 E}{5g_0 + g_2} - 2 (\alpha^2(x) + \beta^2(x)})$,}
\end{align}
\end{subequations}
\end{widetext}
where $f(x)$ is $\Sech(x)$ for bright solitons and $\Tanh(x)$ for dark solitons, also we have omitted the $x$ dependence on $\alpha$, $\beta$, and $\gamma$ in the when using ket notation.

\begin{figure}[!h!t!b!p]
\centering
\includegraphics[width=.5\linewidth]{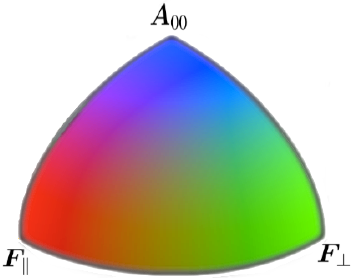} 
\caption{\footnotesize  The color map that is used for the plots in the one dimensional plots section of this letter.  The exact map is given in  (\ref{colormap}).  The color black is reserved for solitons with $|\bm{F}_\parallel| = |\bm{F}_\bot| = |\bm{A_{00}}| = 0$. }\label{fig:colorlegend}
\end{figure}

To further classify these solutions, we consider the value of the spin and singlet terms for each solution.  Since, the solutions are written in the $z$-basis we classify the various solutions in terms of their spin parallel to the $z$-axis $\bm{F}_\parallel$, the spin perpendicular to the $z$-axis $\bm{F}_\bot$, and the singlet term $\bm{A}_{00}$, which  are respectively given by
\begin{subequations}
\label{cmap}
\begin{align}
\bm{F}_\parallel &= \left(\psi_1^2- \psi_{-1}^2\right)+2 \left(\psi_2^2-\psi_{-2}^2\right) \\ 
\bm{F}_\bot &= \sqrt{|\bm{F}_{x}|^2 + |\bm{F}_{y}|^2} \hspace{.4 cm} \text{and,}\\ 
 \bm{A}_{00} &= \frac{\psi_0 \psi_0^* - \psi_1 \psi_{-1}^* - \psi_{-1} \psi_{1}^*+ \psi_{2}\psi_{-2}^* +  \psi_{-2}\psi_{2}^*}{\sqrt{5}}
,
\end{align}
\end{subequations}
where 
\begin{subequations}
\label{fplusminus}
\begin{align}
\nonumber
\bm{F}_{x} &= (\psi_{2}^*\psi_{1} + \psi_{-1}^*\psi_{-2} + \psi_{2}\psi_{1}^* + \psi_{-1}\psi_{-2}^*) \\ 
& + \sqrt{\frac{3}{2}}(\psi_{1}^*\psi_{0} + \psi_{0}^*\psi_{-1} + \psi_{1}\psi_{0}^* + \psi_{0}\psi_{-1}^*) 
\\
 \nonumber
\bm{F}_{y} &=  i (\psi_{2}\psi_{1}^* + \psi_{-1}\psi_{-2}^* - \psi_{2}^*\psi_{1} - \psi_{-1}^*\psi_{1}) \\ 
& - \sqrt{\frac{3}{2}}(\psi_{1}^*\psi_{0} + \psi_{0}^*\psi_{-1} - \psi_{1}\psi_{0}^* - \psi_{0}\psi_{-1}^*).
\end{align}
\end{subequations}
\begin{table}[!h!t!b!p]
\scriptsize
\begin{center}
\begin{tabular}{|c|c|c|c|}
\hline
\multicolumn{4}{|c|}{Solution Characteristics} \\
\hline
 & $|\bm{F}_\parallel|$ & $|\bm{F}_\bot|$ & $|\bm{A_{00}}|$\\
\hline 
1 & $\frac{2 E }{g_0+4 g_1}$ & 0 & 0  \\[.2cm]
2 & $\frac{2 E }{g_0+4 g_1}$ & 0 & 0 \\[.2cm]
3 & $\frac{E }{g_0+g_1}$ & 0 & 0  \\[.2cm]
4 & $\frac{E }{g_0+g_1}$ & 0 & 0  \\[.2cm]
5 & 0 &  0 & $\frac{\sqrt{5} E }{5 g_0+g_2}$ \\[.2cm]
6 & 0 & 0 &  $\frac{\sqrt{5} E }{5 g_0+g_2}$ \\[.2cm]
7 & 0 & 0 &  $\frac{\sqrt{5} E }{5 g_0+g_2}$ \\[.2cm]
8  &  0 & 0 & 0 \\[.2cm]
9  &  0 & 0 & 0 \\[.2cm]
10 & 0 & 0 &  0 \\[.2cm]
11 & 0 & 0 & $\frac{\sqrt{5} E }{5 g_0+g_2}$  \\[.2cm]
12 & 0 & $2 c_1 \sqrt{\frac{30 E -12 c_1^2 \left(5 g_0+g_2\right)}{5 g_0+g_2}}$ & $\frac{\sqrt{5} E }{5 g_0+g_2}$  \\[.2cm]
13 & 0 &  $\frac{E }{g_0+g_1}$ & 0  \\[.2cm]
14 & 0 & $\frac{\sqrt{3} E }{2 g_0}$& 0 \\[.2cm]
15 & 0 & $\frac{E}{2 g_0+8 g_1}$ & 0\\[.2cm]
16 & 0 & $2 c_1  \sqrt{\frac{30 E -12 \left(c_1^2+c_2^2\right) \left(5 g_0+g_2\right)}{5 g_0+ g_2}}$ & $\frac{\sqrt{5} E }{5 g_0+g_2}$\\[.2cm]
\hline
\end{tabular}
\end{center}
\caption{\footnotesize  The characteristics of the Thomas-Fermi one dimensional solutions. Note that since the coefficients are real valued by design, $\bm{F}_{y} = 0$ and $\bm{F}_\bot = \bm{F}_{x}$.   Only solution nos. 12 and 16 show hybridization between the solution characteristics a the spin-2 superfluid.}
\label{tab:Sol-Chars}
\end{table}
Assigning each of these characteristics a primary color can be a useful aid when visualizing the dynamic evolution of spin-2 solitons.  We implement a direct color map where $\bm{F}_\parallel$, $\bm{F}_\bot$, $\bm{A}_{00}$ directly to red, green, and blue colors respectively.  This will become quite useful in the analysis of soliton collisions. The exact mapping is given by  
\begin{align} \label{colormap}
\frac{\left(|\bm{F}_\parallel|,\text{ } |\bm{F}_\bot|, \text{ }|\bm{A_{00}}|\right)}{\sqrt{|\bm{F}_\parallel|^2 + |\bm{F}_\bot|^2 + |\bm{A_{00}}|^2}} \to (r, g, b) . 
\end{align}
The normalization guarantees that the color lies in the first octant a $(r, g, b)$ color sphere as shown in Fig.~\ref{fig:colorlegend}.  For the case where $|\bm{F}_\parallel| = |\bm{F}_\bot| = |\bm{A_{00}}| = 0$ we reserve the color black.
\subsection{ Energy eigenstate solutions} 
The spin-2 BEC equation of motion is a system of five non-linear differential equations given by (\ref{eomspin2}). This set of equations does not currently have a known general solution.  In the one-dimensional case, it is possible to find several different multichannel (multiple $m_f$ levels excited) exact energy eigenstates solutions.  The general technique used is to reduce the system of five nonlinear partial differential equations to a system of five nonlinear algebraic equations.  The inspiration for this technique was the work of \cite{zhang:2012}.
\subsubsection{1 dimensional energy eigenstates}
For a wave function $\Psi(x, t)$ to be an energy eigenstate of the spin-2 BEC equation it must satisfy 
\begin{align}
\label{energyeigenstate}
\frac{i \partial \Psi_m(x,t)}{\partial t}= E \Psi_m(x, t),
\end{align}
where $E$ is the energy eigenvalue.  The form of the solutions we will looks for will be solutions where the space and time variables are separated specifically solutions of the form
\begin{align}\label{timeseparation}
\Psi_m(x,t) = \psi_m(x)e^{-i E t}.
\end{align}
Thus taking the derivative on the left hand side of (\ref{eomspin2}) yields 
\begin{align}\label{eigeneomspin2}
E \psi_m(x) &= \left(-\partial_{xx} + g_0 \left |\psi(x) \right |^2 \right)\psi_m(x) \nonumber \\
&+ g_1 \sum_{m' = -2}^{m'=2}\bm{F} \cdot \bm{f}_{mm'}\psi_m(x) + g_2 \left|{A}_{00}\right|^2\psi^*_{-m}(x). 
\end{align}
Notice that the interaction terms on right hand side of  (\ref{eigeneomspin2}) are all of the form $\psi_m(x) \psi_{m'}(x) \psi^*_{m''}(x)$.  So, if we make the assumption that $\psi_{m}(x) = c_m f(x)$, where $f(x)$ is a real valued function, and $c_m \in \mathbb{R}$ we can rearrange the equation of motion to be 
\begin{align}\label{simpleeom}
\partial_{xx} \left(c_mf(x)\right) = E \psi_m(x) + \sum c_mc_{m'}c_{m''}f(x)^3.
\end{align}
In  (\ref{simpleeom}) the indices in the sum are suppressed since the focus should be on finding a function $f(x)$ such that 
\begin{align}
\partial_{xx}\left(f(x)\right) = a f(x) + b f(x)^3,
\end{align} 
where $a$ and $b$ are real numbers.  There are two elementary functions that meet this requirement, $\Sech(k x)$ and $\Tanh(k x)$, which will produce bright and dark soliton solutions respectively.
\subsubsection{Bright soliton solutions}
For a bright soliton solution we assume 
\begin{align} 
\psi_{m}(x) = c_{m} k \Sech(k x),
\end{align} where we choose $k$ such that  ${E} = -k^2$.  Plugging in our $\psi_m(x)$ into  (\ref{eigeneomspin2}) gives  
\begin{align}
\label{system_of_equations_bright}
\begin{pmatrix}
0\\
0\\
0\\
0\\
0\\
\end{pmatrix}
= k^3 \Sech^3(kx) \begin{pmatrix}
2 c_2  + \rho c_2g_0 + \xi_2 g_1 + |A_{00}| c_{-2}g_2\\
2 c_1 + \rho c_1g_0 + \xi_1 g_1 + |A_{00}| c_{-1}g_2\\
2 c_0 + \rho c_0g_0 + \xi_0 g_1 + |A_{00}| c_{0}g_2\\
2 c_{-1} + \rho c_{-1}g_0 + \xi_{-1} g_1 + |A_{00}| c_{1}g_2\\
2 c_{-2} + \rho c_{-2}g_0 + \xi_{-2} g_1 + |A_{00}| c_{2}g_2\\
\end{pmatrix},
\end{align}
where $\alpha$, $\beta$, and $\xi_{m_f}$ are the same as in  (\ref{alphabeta}).  This system of equations is the exact same set of equations that came from the Thomas-Fermi approximation as written in  (\ref{system_of_equations}) multiplied by $k^3\Sech^3(k x)$, and with $E = -2$.  Hence, the coefficients for the bright soliton solutions are the same as the Thomas-Fermi coefficients with $E$ set to $-2$.  It is worth noting that in the region where the Thomas-Fermi approximation applies, at $x \to \pm \infty$, all the bright soliton solutions approach the trivial Thomas-Fermi solution where all $c_m = 0$ since $\Sech(x) \to 0$ as $x \to \pm \infty$.   
\subsubsection {Bright soliton energy and momentum} 
The energy of the spin-2 BEC quantum fluid is given by the matrix element of the time deriviative
\begin{equation}
E = \bra{\psi}i \partial_{t}\ket{\psi},
\end{equation} 
which for a stationary bright soliton solution of form (\ref{timeseparation})  is $E=-k^2 \bra{\psi}\ket{\psi}$. This reduces to $E=-k^2$ when $\psi$ is normalized ($\bra{\psi}\ket{\psi}=1$), but we have left the $\bra{\psi}\ket{\psi}$ since the solutions we have given are not inherently normalized.  This will remain the convention throughout this section.  To give a quantum fluid a momentum $p$ in the x direction we simply multiply the field $\psi$ by $e^{- i p x}$ since the momentum is given by 
\begin{equation} \label{momentum_p}
\bra{\psi(x)}- i \partial_{x}\ket{\psi(x)} = p \bra{\psi}\ket{\psi} - i \int{\psi^*(x) \psi ' (x)  dx},
\end{equation} 
where the integral vanishes for even or odd functions of a constant phase.  The time dependent solution for a  bright soliton with momentum $p$, remember $m = 1/2$, is given by 
\begin{equation} \label{timedependentsoliton}
\Psi_m (x, t) = k c_m e^{i p x} e^{i \left(k^2-p^2\right) t} \text{sech}(k (x-2 p t)).
\end{equation}
This gives a dispersion relation for a bright soliton with momentum $p$ of 
\begin{subequations}
\label{dispersion}
\begin{align}
E &= \bra{\psi}i \partial_{t}\ket{\psi} \\ 
& = (p^2 - k^2 )\bra{\psi}\ket{\psi}  + \mathbb{Z}  \\ 
& =   (p^2 - k^2) \bra{\psi}\ket{\psi},
\end{align}  
\end{subequations}
where 
\begin{equation}
\mathbb{Z} = 2 i k p\int{\small \tanh (k (x - 2 p t)) \sech(k (x - 2 p t)) dx} = 0.
\end{equation}
Hence the dispersion relation for bright solitons is given by 
\begin{align}
   E = p^2 - k^2. 
   \end{align}
 If a chemical potential term is included in the equation of motion the dispersion relation is modified to  
 \begin{align}
E = \mu + p^2 - k^2.
   \end{align}
These solution are included in Table~\ref{Table_I_1D_solutions}. 

\subsubsection{Dark soliton solutions} \label{kinkstates}   
Alternatively to the bright soliton solutions presented in the previous section, one can assume a solution of the form  
\begin{align}
\Psi_{m} (x, t) = c_{m} k \Tanh(k x) e^{i E t}, 
\end{align}
where now 
\begin{align}
k = \sqrt{\frac{E}{2}} .  
\end{align}
This will give a dark soliton solution.  This results in a system of equations that is identical to (\ref{system_of_equations}) multiplied by $k^3\Tanh^3(k x)$ and with $E = 2$.  Specifically, we get 
\begin{align}
\label{system_of_equations_kinks}
\begin{pmatrix}
0\\
0\\
0\\
0\\
0\\
\end{pmatrix}
= k^3 \Tanh^3(kx) \begin{pmatrix}
-2 c_2  + \rho c_2g_0 + \xi_2 g_1 + |A_{00}| c_{-2}g_2\\
-2 c_1 + \rho c_1g_0 + \xi_1 g_1 + |A_{00}| c_{-1}g_2\\
-2 c_0 + \rho c_0g_0 + \xi_0 g_1 + |A_{00}| c_{0}g_2\\
-2 c_{-1} + \rho c_{-1}g_0 + \xi_{-1} g_1 + |A_{00}| c_{1}g_2\\
-2 c_{-2} + \rho c_{-2}g_0 + \xi_{-2} g_1 + |A_{00}| c_{2}g_2\\
\end{pmatrix},
\end{align}
where $\alpha$ and $\beta$ are once again the same as in  (\ref{alphabeta}).  Once again we need to make assumptions on the $c_m$ to solve the system of equations.  It should come as no surprise that the solutions for the dark soliton coefficients are identical to the solutions for the Thomas-Fermi solutions with $E \to 2$.   In this case as $r \to \infty$ almost every dark soliton solution approaches it's corresponding Thomas-Fermi  solution in the limit since 
\begin{align}
\lim_{x \to \infty} c_{m} k \Tanh(k x) = c_{m} k 
& = c_{m} \sqrt{\frac{E}{2}},
\end{align} 
which is equal to the Thomas-Fermi solution coefficients if all $c_m$ are proportional to $\sqrt{E}$.  This is the case for all solutions except for solutions nos. $11$, $12$, and $16$.  For solution nos. $11$, $12$, and $16$, $c_1$ and $c_2$ must take on specific values to match the Thomas-Fermi background field limit.  Specifically,
\begin{align}
c_1 = c_2 = \sqrt{E}.
\end{align}
These solution are included in Table~\ref{Table_I_1D_solutions}. 

\subsubsection {Dark soliton energy and momentum} 
The time dependent solution for a dark soliton moving at a momentum $p$ is 
\begin{equation} \label{timedependentdarksoliton}
\Psi_m (x, t) = k c_m e^{i p x} e^{-i \left(2 k^2+p^2\right) t} \text{tanh}(k (x-2 p t)).
\end{equation}
Calculating the dispersion relation for (\ref{timedependentdarksoliton}) gives 
\begin{subequations}
\label{darkdispersion}
\begin{align}
E &= \bra{\psi}i \partial_{t}\ket{\psi} \\ 
& = (2 k^2 + p^2)\bra{\psi}\ket{\psi}  - \mathbb{Z}  \\ 
& =   (2 k^2 + p^2) \bra{\psi}\ket{\psi},
\end{align}  
\end{subequations}
where 
\begin{equation}
\mathbb{Z} = 2 i k p\int{\small \tanh (k (x - 2 p t)) \sech ^2 (k (x - 2 p t)) dx} = 0.
\end{equation}
This leads to a parabolic  dispersion relation $E = p^2 + 2 k^2$, or $E = \mu + p^2 + 2 k^2$ if the equations of motion include a chemical potential. Note that the energy for the dark solitons differs only in the sign and coefficient of the $k^2$ term.

\section{Quantum simulations in 1D}
\label{Section_Quantum_simulations_in_1D}

There are two main categories of experiments we perform: (1) calibration experiments and (2) scattering experiments.  In the calibration experiments  we initialize the quantum lattice gas algorithm with an energy eigenstate solution and evolve it in time for a very long time to make sure that the wave function is indeed stationary.  Once we have a full suite of calibration of experiments, we can run more exploratory experiments where we can view the BEC quantum fluid reacting to novel circumstances and nonintegrable initial conditions. An example of a one dimensional scattering experiment  is a bright soliton collision experiment, where we take one bright soliton and give it an initial fixed non-zero momentum so that it can subsequently collide with a different stationary bright soliton.

\subsection{Calibration (stationary) quantum simulations}  

\subsubsection{Bright solitons}

\begin{figure}[!h!t!b!p]
  \centering
  \includegraphics[width= 1\linewidth]{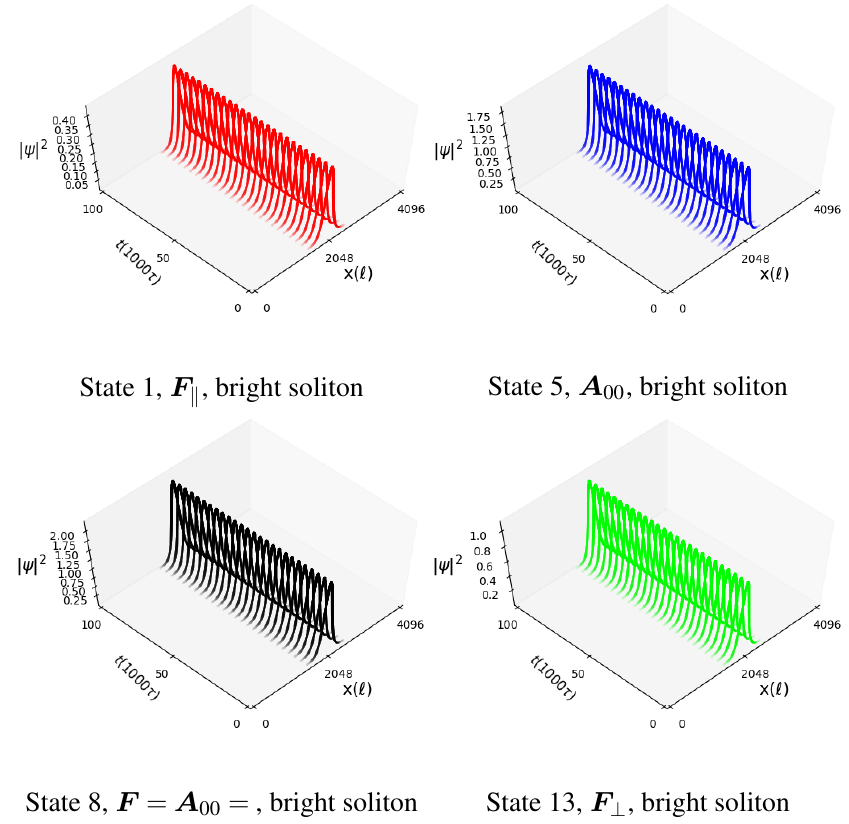} 
  \caption{  \label{fig:scattering1_bright} \footnotesize Four bright solitons each one with a different characteristic color as described in  (\ref{colormap}) and Fig.~\ref{fig:colorlegend}.  The plot shows that the solutions are stationary for 100,000 time steps.}
\end{figure}

Presented here  is an example waterfall plot demonstrating the temporal stability of the bright soliton energy eigenstates---see Fig.~\ref{fig:scattering1_bright}.  Notice how the density of the wave packet does not change with time, as was predicted analytically. 

\subsubsection{Dark solitons}
\begin{figure}[!h!t!b!p]
  \centering
    \includegraphics[width= 1\linewidth]{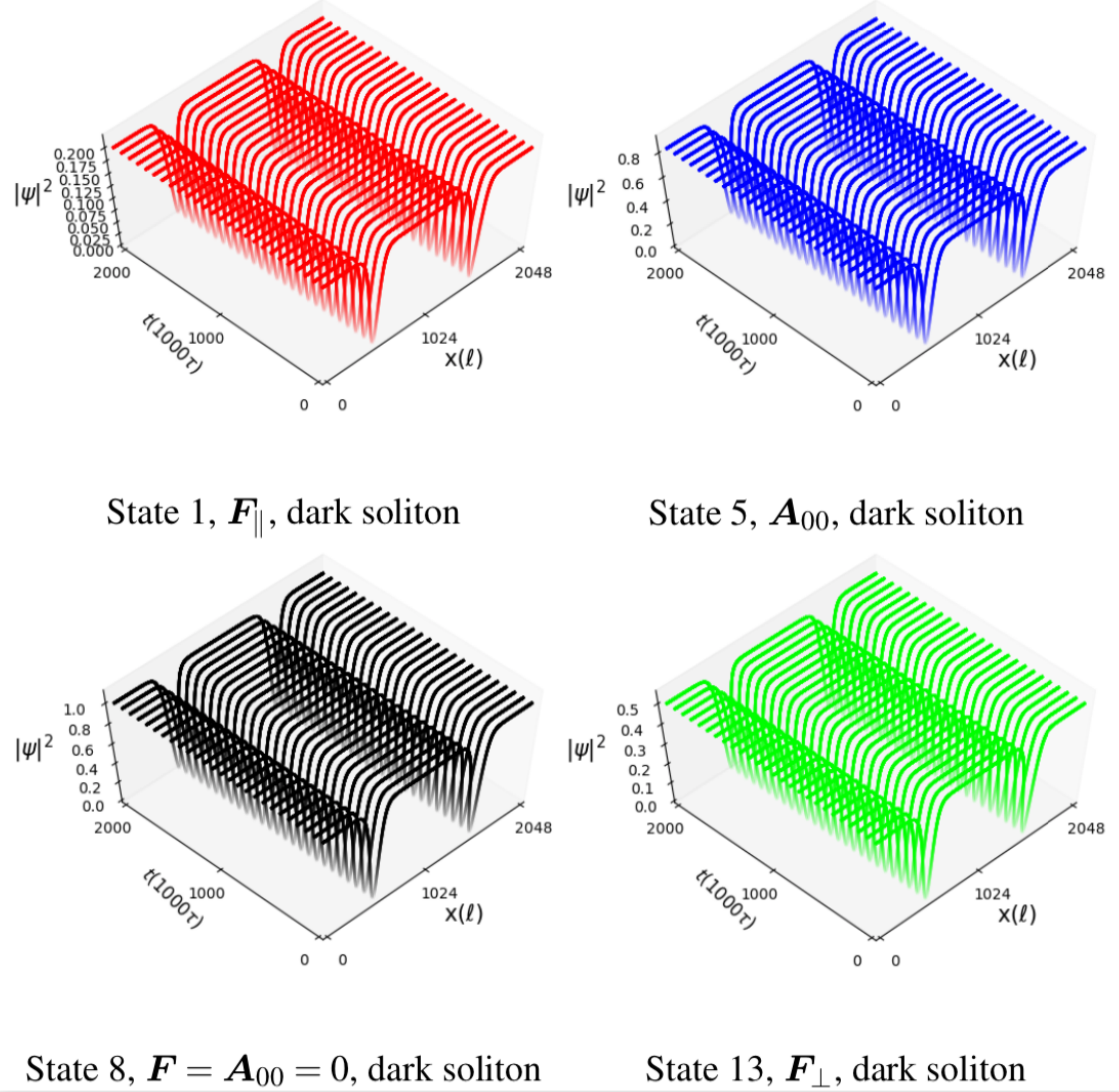} 
  \caption{  \label{fig:scattering1_dark}
\footnotesize Four dark solitons each one with a different characteristic color as described in  (\ref{colormap}) and Fig.~\ref{fig:colorlegend}. The plot shows that the solutions are stationary for 200,000 time steps.}
\end{figure}

The dark soliton solutions are also stationary as is predicted analytically---see Fig.~\ref{fig:scattering1_dark}.  However, since the dark solitons are proportional to $\Tanh(x)$ and since the hyperbolic tangent function approaches one at infinity and negative one at negative infinity, we must include two copies of the solution to ensure periodic boundary conditions.  To create a smooth transition from one soliton to the next as well as a smooth transition across the boundary we simply multiply two dark soliton solutions together making the total field%
\begin{align}
\psi_{total}(x) = \psi(x - \tfrac{L}{4})\psi(\tfrac{3L}{4} - x)
,
\end{align}
where $L$ is the total length of the lattice.

\subsection{Kinetic experiments}
\subsubsection{Kicked solitons}
\begin{figure}[!h!t!b!p]
{\xy(0, 0)*{\includegraphics[width=.65\linewidth]{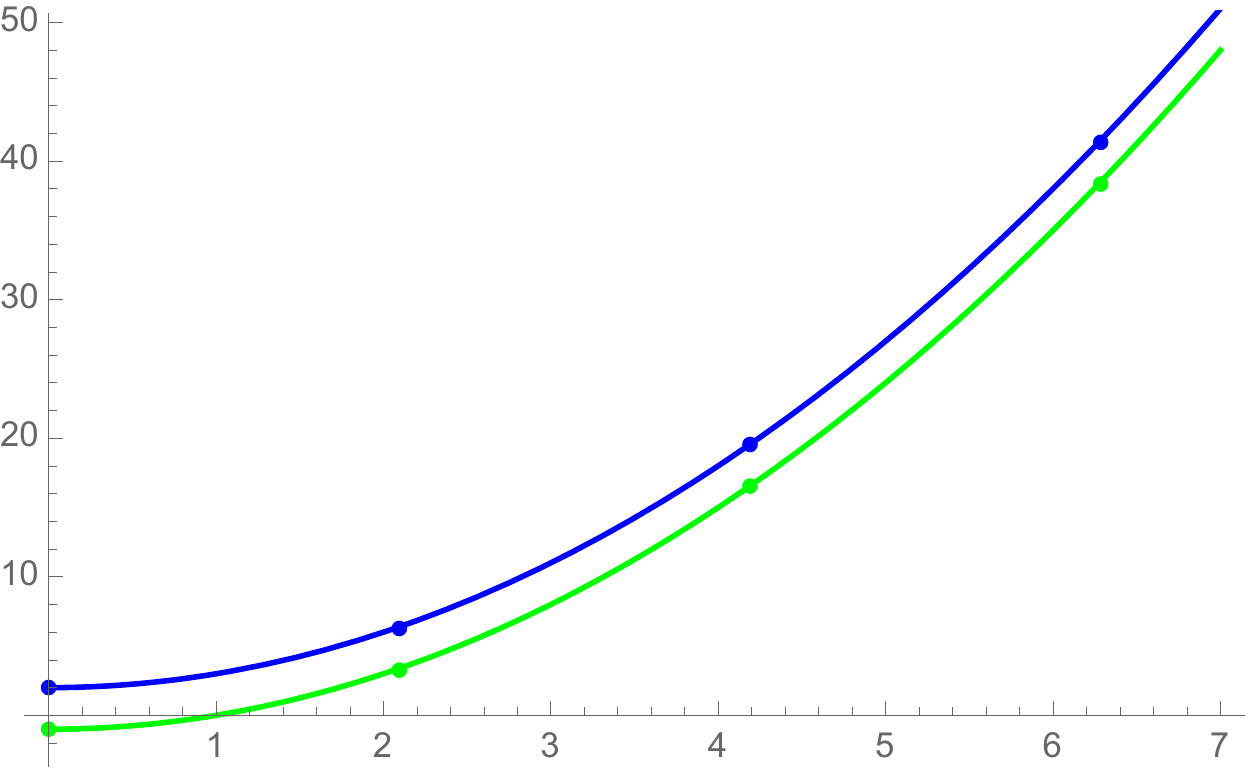}};
(-35, 0)*{\text{\scriptsize{$E\; \bigl(\frac {1}{\ell^2}\bigr)$}}};
(0, -20)*{\text{\scriptsize{$p\; \bigl(\frac {1}{\ell}\bigr)$}}};
\endxy}
\caption{\footnotesize  The energy plotted as a function of momentum for bright (green) and dark (blue) solitons both with $k = 1$.  The lines are the theoretically predicted energies and the markers are the energies calculated using the evolution of the simulation.}
  \label{fig:dispersions}
\end{figure}

Now that we are confident   the energy eigenstates are indeed stationary when using the quantum lattice gas algorithm, we can give these soliton solutions momentum by multiplying the quantum field by $e^{ i {2 \pi n x}/{L}}$, where $n$ is an integer to ensure periodic boundary conditions.  This gives the solitons a momentum $p = {2 \pi n }/{L}$.
\begin{figure}[!h!t!b!p]
  \centering
  \includegraphics[width= 1\linewidth]{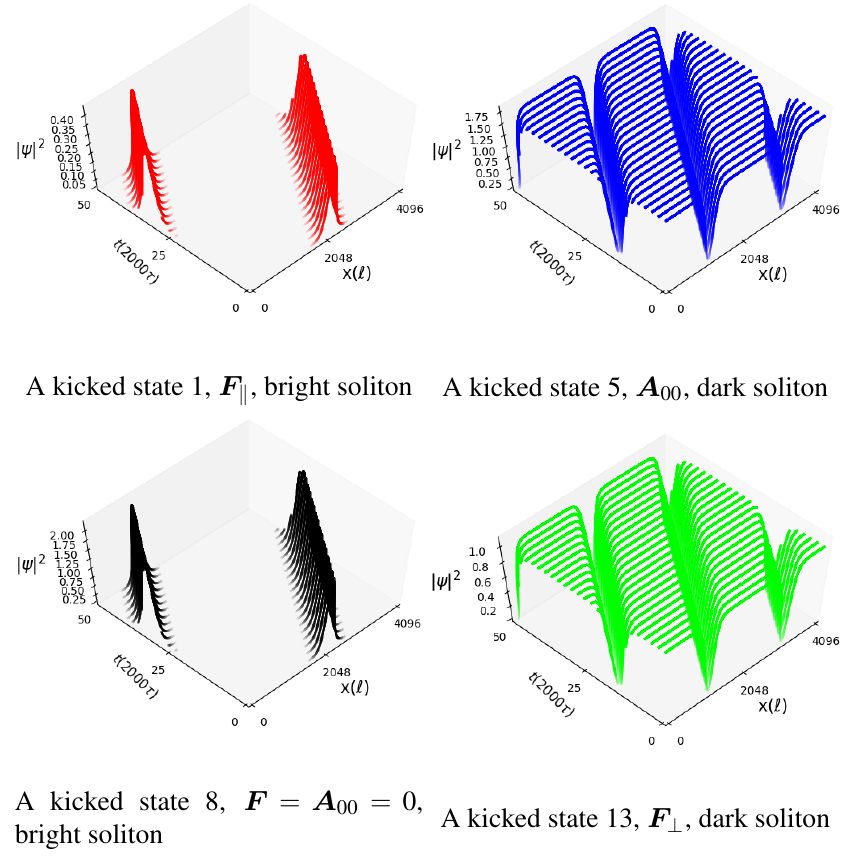} 
  \caption{\footnotesize Kicked bright and dark solitons each with a different characteristic color as described in  (\ref{colormap}) and Fig.~\ref{fig:colorlegend}. The plot shows that both bright and dark soliton solutions move together in packets for 100,000 time steps.}
\label{fig:kicked solitons}
\end{figure}
Notice how in Fig.~\ref{fig:kicked solitons} the solitons move together in a particle like fashion.  Now that we are able to give the particles a momentum we can respectively test the dispersion relations in (\ref{dispersion}) and (\ref{darkdispersion}) with the results of our simulations.  A discrete equation to calculate the energy for our simulations is given by 
\begin{equation}\label{disecrete_energy}
E = i \frac{\bra{\psi(t)}\ket{\psi(t)} - \bra{\psi(t)}\ket{\psi(t-\tau)}}{\bra{\psi(t)}\ket{\psi(t)}}.
\end{equation}  
Fig.~\ref{fig:dispersions} shows a nearly identical energy dispersion between the simulation and the theoretical predictions. 
\subsubsection{Bright soliton breathers}
A slightly perturbed bright soliton solution can also become a breather soliton. Also, if the perturbation is too big one can split a multiple channel soliton into breather solitons in its different components.  A breather soliton is a soliton that oscillates in periodically about the soliton solution.  The periods of these breathers as well as the maximum size of the perturbation they can withstand the nonlinear interaction before they begin to breakdown are both potential areas of study \cite{trombetoni, Golde}.  For our purposes we are content to simply observe that the quantum lattice gas algorithm can reproduce both of these phenomenological features. 
\begin{figure}[!h!t!b!p]
  \centering
    \includegraphics[width= 1\linewidth]{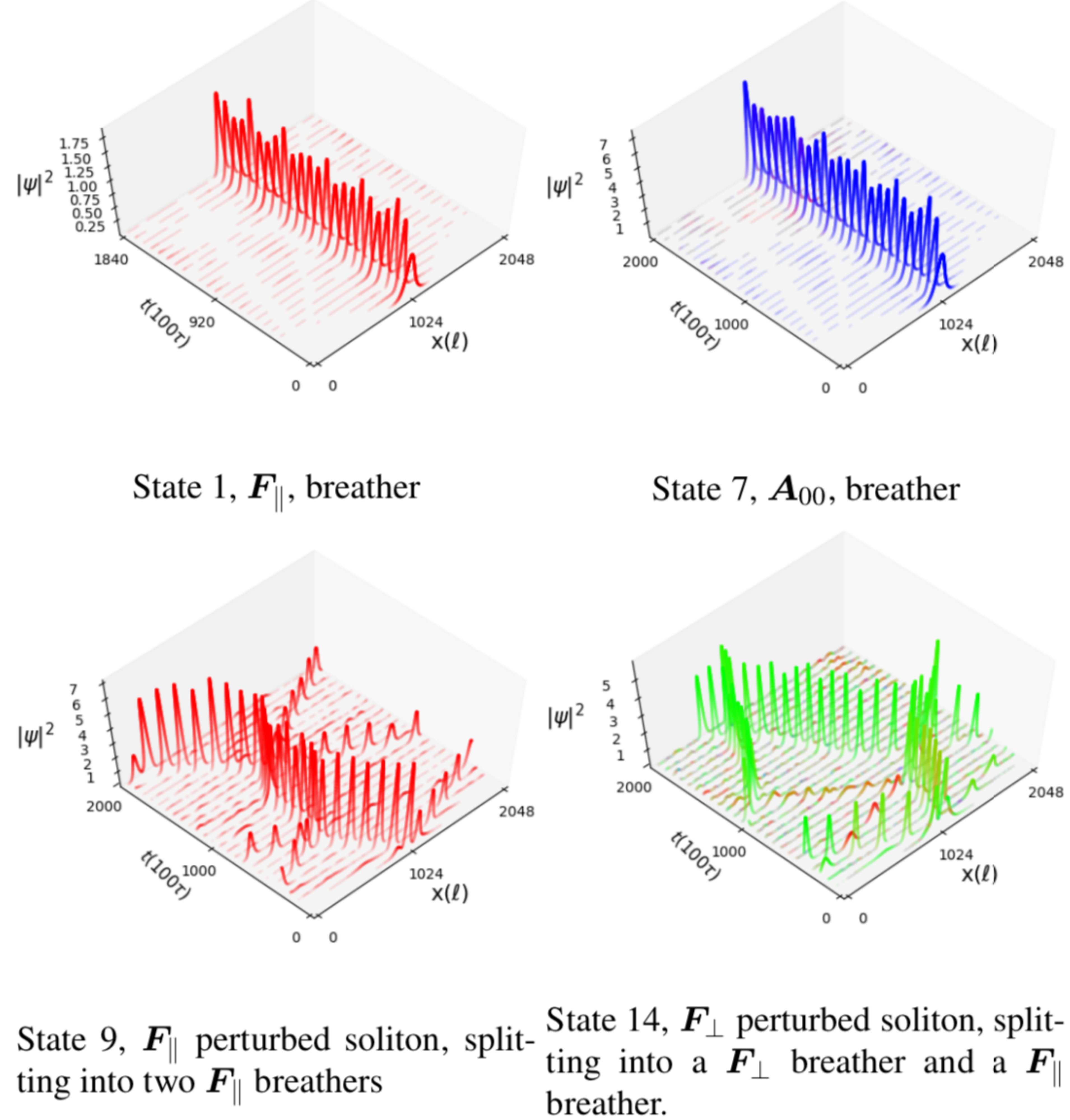} 
  \caption{\footnotesize Four breather solitons each where the color represents the spin characteristics as described in  (\ref{colormap}) and Fig.~\ref{fig:colorlegend}.  The solitons are initialized with the double the coefficients for their respective stationary solitons.  For state 1 and state 7 the spin-2 BEC simply oscillated about its initial configuration, while for states 9 and 14 the BEC splits into two separate breathers.}
  \label{fig:breathers 1}
\end{figure}
\subsubsection{Bright soliton collision}
\begin{figure}[!h!t!b!p]
  \centering
      \includegraphics[width= 0.9\linewidth]{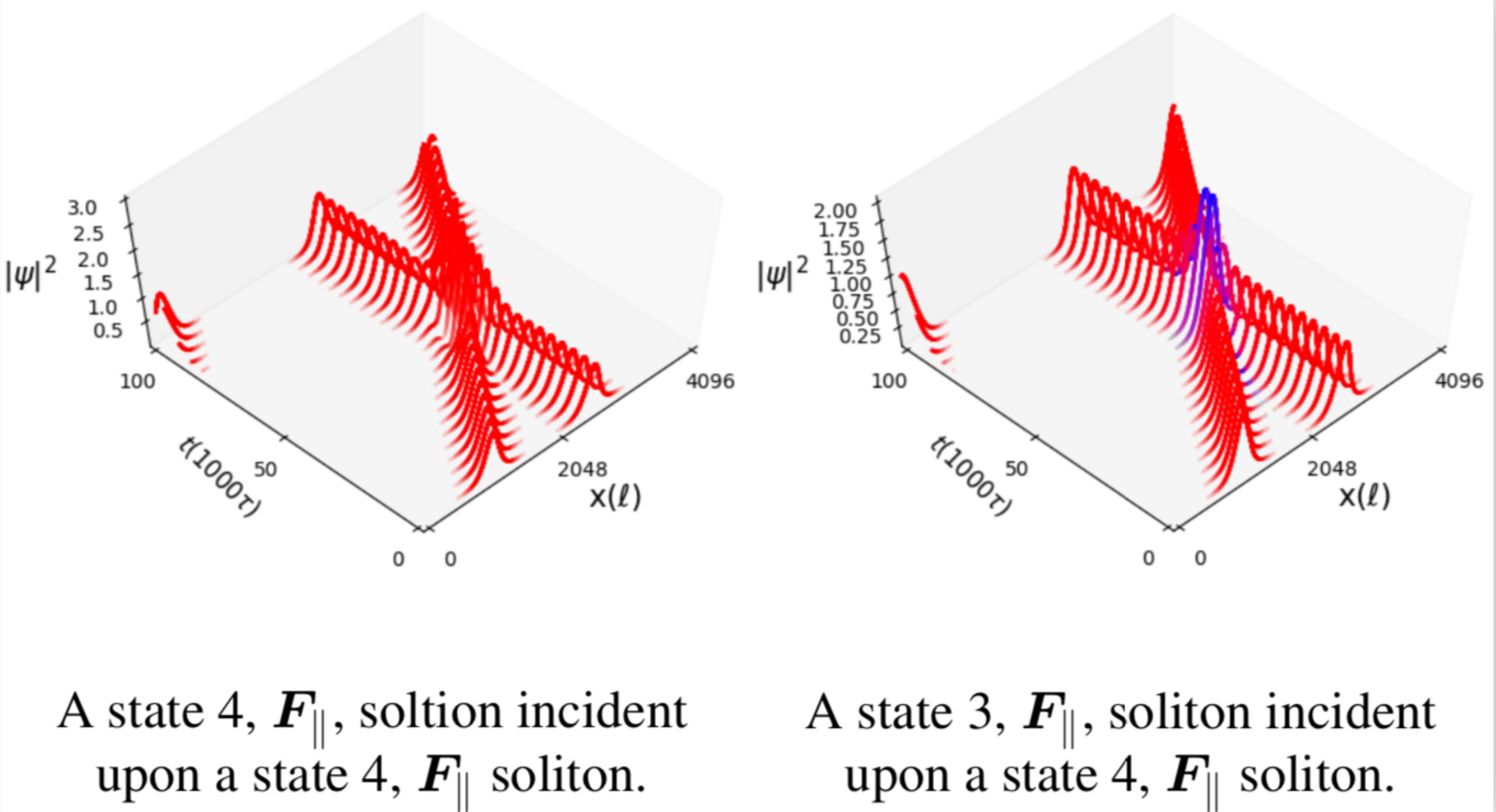} 
  \caption{  \label{fig:scattering1}\footnotesize  In the collision on the left we see two solitons passing through each other.  Whereas in the collision on the right where the solitons interact via the $\bm{A}_{00}$ singlet where they overlap.}
\end{figure}

\begin{figure}[!h!t!b!p]
  \centering
       \includegraphics[width= 0.9\linewidth]{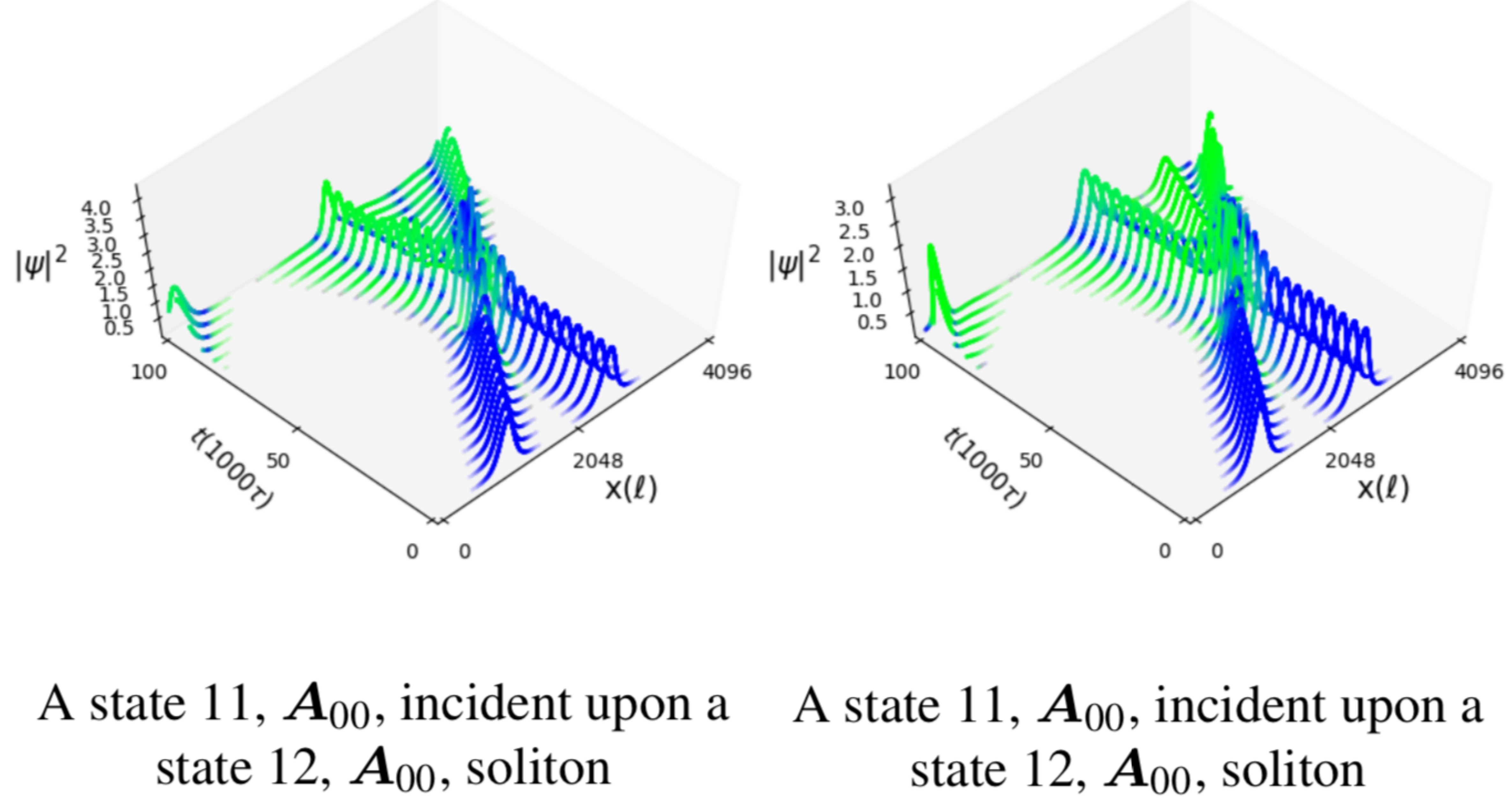} 
  \caption{\footnotesize  In the collision on the left we see two solitons dominated by the $\bm{A}_{00}$ interaction until they collide at which point the solitons become bound by the  $\bm{F}_{\parallel}$ interaction.  Whereas in the collision on the right where a similar interaction occurs with an additional $\bm{F}_{\bot}$ soliton trailing the outgoing soliton.}
  \label{fig:scattering2}
\end{figure}

\begin{figure}[!h!t!b!p]
  \centering
       \includegraphics[width= 0.9\linewidth]{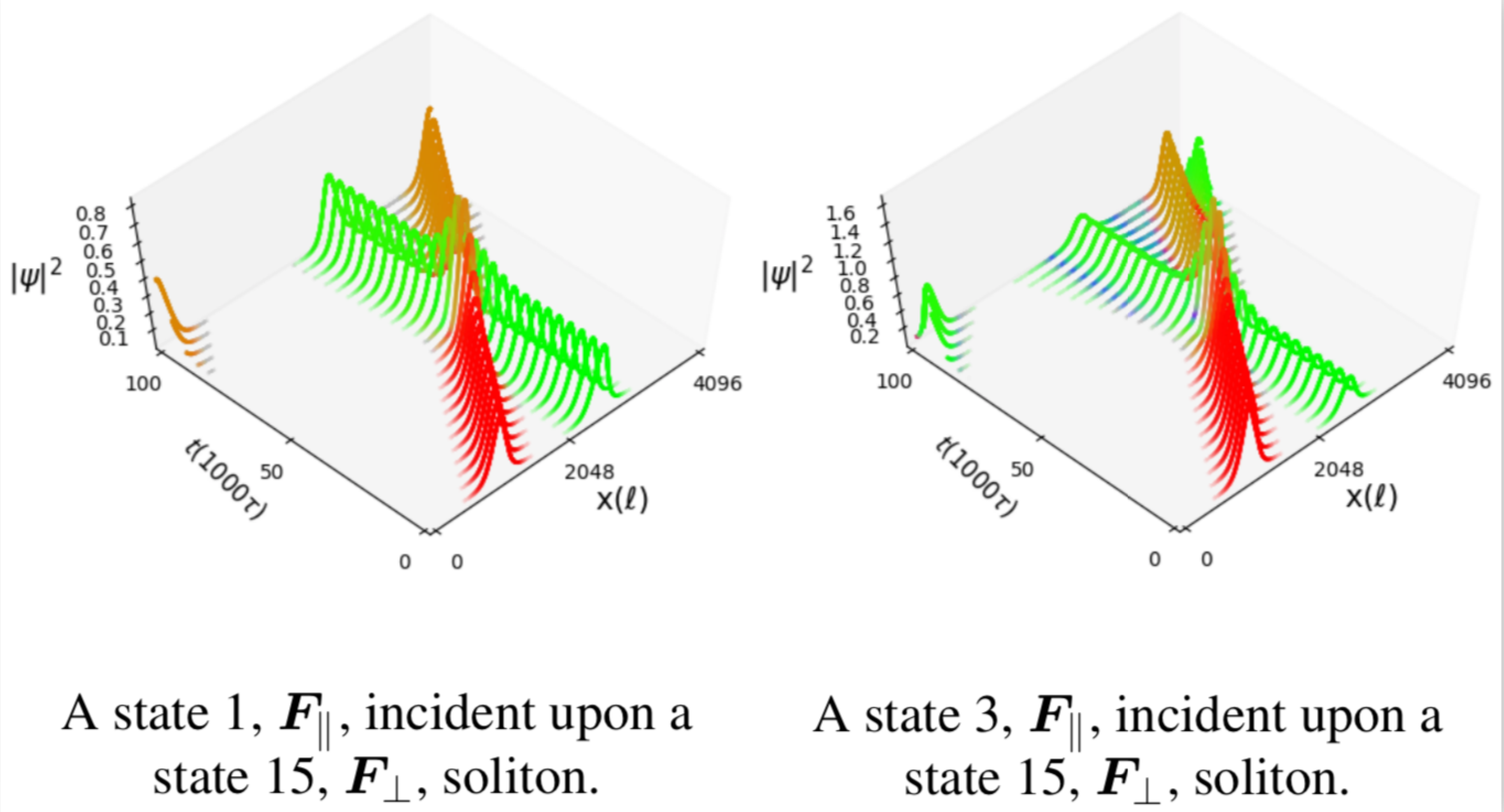} 
\caption{\footnotesize  In the collision on the left we see two solitons passing through each other where incoming soliton goes from being dominated by the $\bm{F}_{\parallel}$ interaction to being bound by a mix of the $\bm{F}_{\parallel}$ and $\bm{F}_{\bot}$ interactions.  Whereas in the collision on the right where we see an almost identical looking interaction but now there is an accompanying $\bm{F}_{\bot}$ soliton that is traveling ahead of the soliton bound by both the  $\bm{F}_{\parallel}$ and $\bm{F}_{\bot}$ interactions which appears gold.}
  \label{fig:scattering3}
\end{figure}

\begin{figure}[!h!t!b!p]
  \centering
         \includegraphics[width= 1\linewidth]{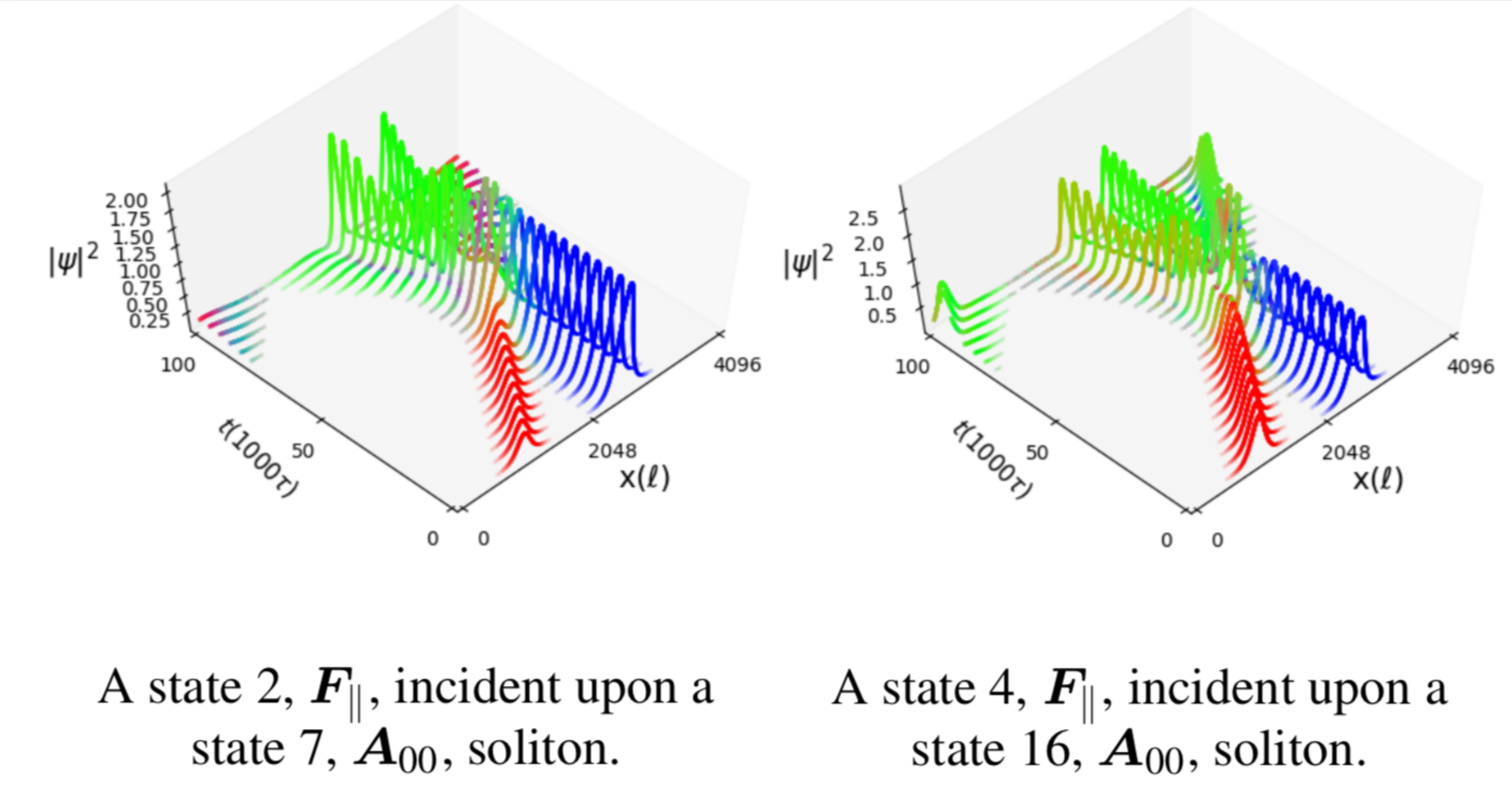} 
  \caption{  \label{fig:scattering4}
  \footnotesize  In the collision on the left we see the incoming $\bm{F}_{\parallel}$ soliton and the stationary collide with a $\bm{A}_{00}$ soliton. After the collision  both solitons become dominated by $\bm{F}_{\bot}$ interaction.  This process appears to slow the incoming soliton down dramatically. Whereas in the collision on the right a similar process occurs however now one of the outgoing solitons also include the $\bm{F}_{\parallel}$ interaction term making a gold color.  There is also a new third soliton that appears to have made it through the stationary soliton without losing velocity. }
\end{figure}

In this section we study what happens when a moving soliton collides into another soliton that is at rest.  The incident soliton with momentum have been given the second lowest possible momentum, meaning that the stationary state was multiplied by the phase $e^{i {4\pi x}/{L}}$. Depending on the two solitons that are colliding we see that  characteristics---such as its color as defined in (\ref{cmap})---of the soliton can changes as a new soliton is formed, thereby preserving the conservation laws.  The following four pairs of soliton solutions were chosen  for the cleanliness of their waterfall plots as well as their similarities to one another.  The idea is that if we narrow our focus onto the simpler interactions it improves our chances at understanding the properties of the interactions on an intuitive level.   Example bright soliton-soliton collisions are presented in Figs.~\ref{fig:scattering1}--\ref{fig:scattering4}.
The soliton collisions are able to change the characteristics of both solitons as well as split the solitons into additional solitons as well as breathers.  There is a huge parameter space to explore given all the different solitons and each collision could behave differently given different relative momenta.  A complete analysis of spin-2 BEC bright soliton collisions remains a topic for future study.

\section{Solutions to the spinor GP equation in 2D}
\label{Section_Solutions_to_the_Spinor_GPE_in_2D}

The solutions to (\ref{eomspin2}) in two-spatial dimensions are more difficult to find analytically since the $\nabla^2$ term is upgraded from $\partial_{xx}$ to $\partial_{xx} + \partial_{yy}$.  Of course the one-dimensional solutions are still solutions in two dimensions, they will just be the same bright or dark soliton solution repeated $y$ times across the lattice.   We are more interested in the intrinsically two dimensional solutions, so far we have have found numerically approximate solutions with radial symmetry using a Pad\'e approximant that is motivated by the $\Tanh(x)$ dark soliton solutions found in 1D.   
\subsection{Single channel Pad\'e dark soliton}
\begin{figure}[!h!t!b!p]
\centering
\begin{tikzpicture}
\node at (0,0) {\includegraphics[width=0.8\linewidth]{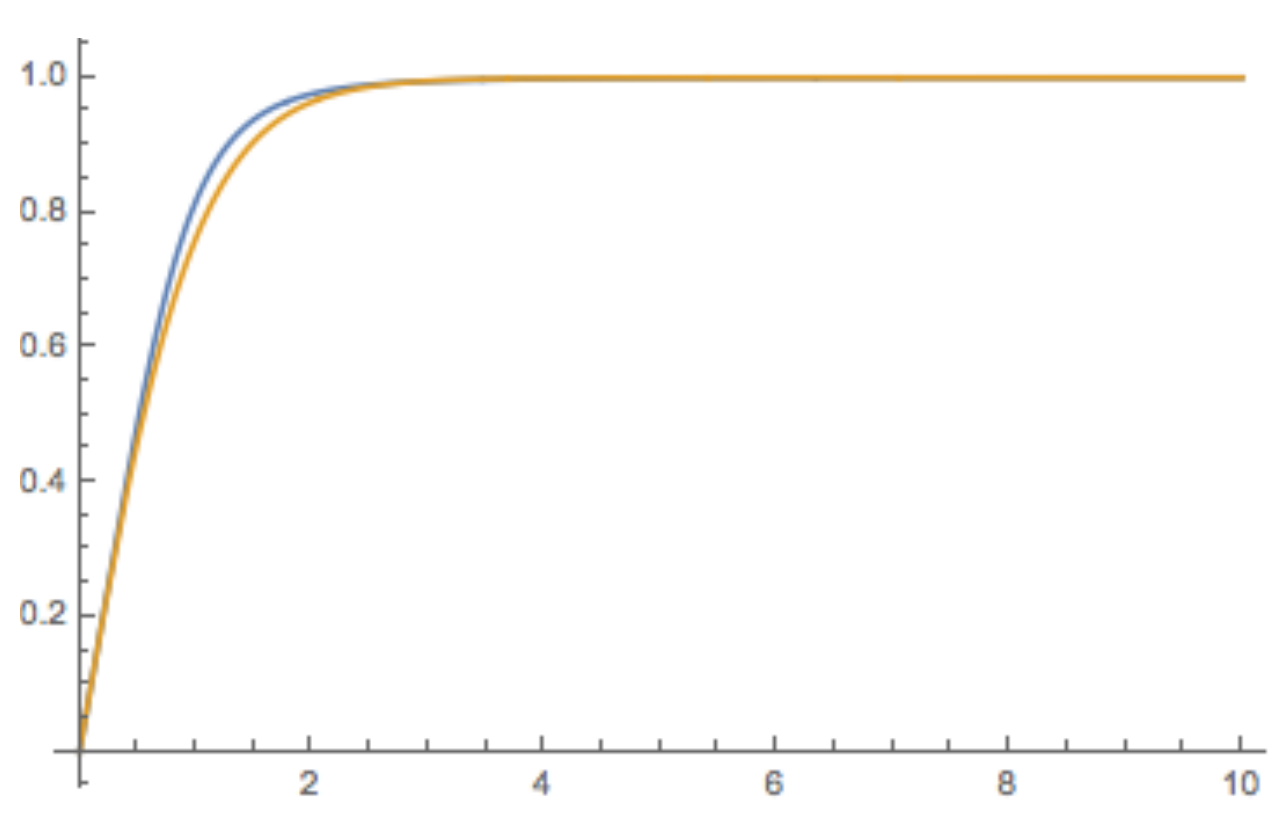}};
\node at (-3.9,0.5) {$|\psi|^2$};
\node at (0,-2.25) {$r\;(\ell)$};
\end{tikzpicture}
\caption{\footnotesize  This is a graph of the Pad\'e approximant given in (\ref{pade_form}) with $a_1$, $a_2$, and $b_1$ all equal to 1 in blue, while the yellow function is $\Tanh(x)$.}\label{fig:pade_approx}
\end{figure}

\begin{figure}[!h!t!b!p]
\centering
\begin{tikzpicture}
\node at (0,0) {\includegraphics[width=0.8\linewidth]{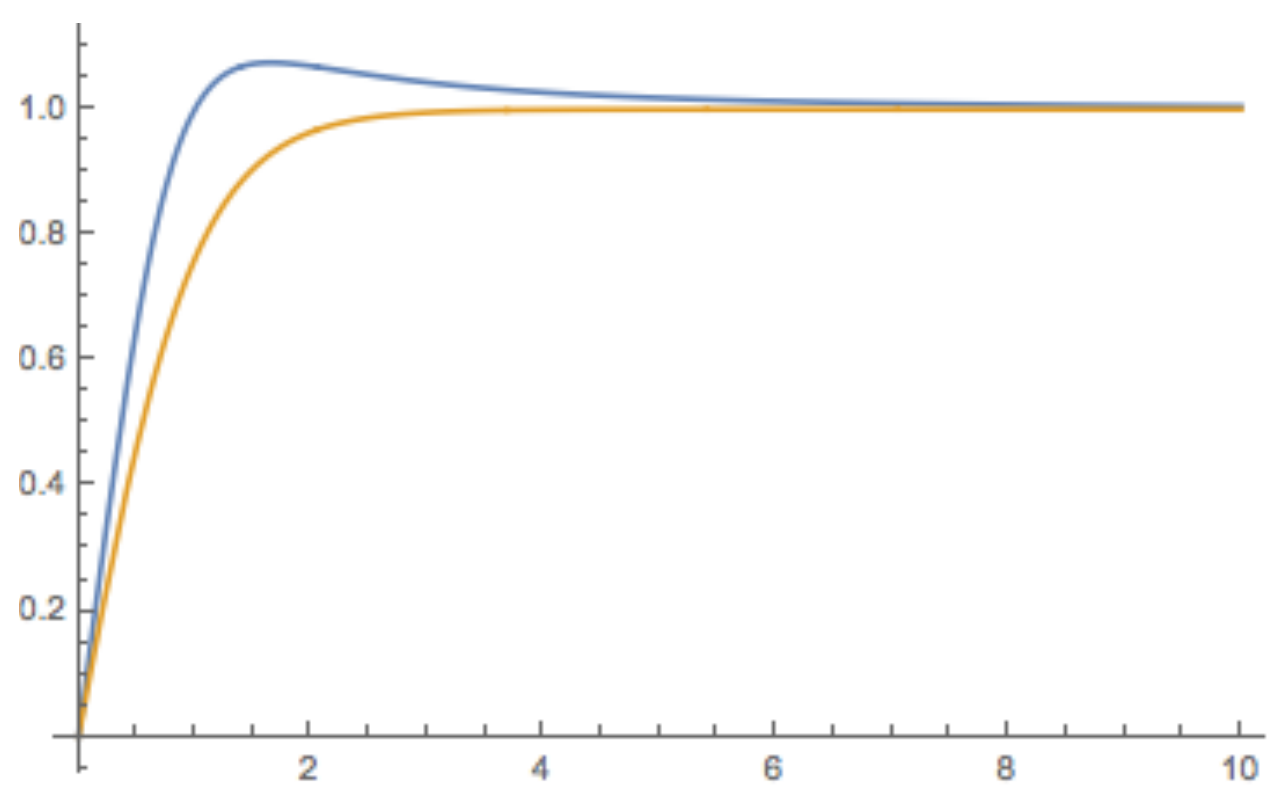}};
\node at (-3.9,0.5) {$|\psi|^2$};
\node at (0,-2.25) {$r\;(\ell)$};
\end{tikzpicture}
\caption{\footnotesize  This is a graph of the Pad\'e approximant given in (\ref{pade_form}) with $a_1$ = 2 and $a_2 = b_1 = 1$ in blue, while the yellow function is once again $\Tanh(x)$.}\label{fig:bad_pade_approx}
\end{figure}

In one dimension, the $\Tanh(x)$ dark soliton solutions have the property that $\Tanh(-x) = -\Tanh(x)$.  Thus to find dark vortex solutions we would want  solutions with the property $\psi(r, \theta) = -\psi(r, \theta + \pi)$.  This is achieved by choosing a Pad\'e approximant that is symmetric in $r$ and multiplying it by $e^{i n \theta}$, where $n$, the winding number, is odd. So far we have only found numerical solutions for $n=1$.  Specifically, the form of our trial solution is    
\begin{align}
\label{pade_form}
\Psi_{m_f}(r, \theta, t) = \sqrt{\frac{a_2 r^4+a_1 r^2}{a_2 r^4+b_1 r^2+1}}e^{i \theta}e^{i E t},  
\end{align}
where $\Psi_{m_f}(r, \theta, t)$ is the wave function in the only occupied level of the spin manifold. This form of Pad\'e approximant was first used by \cite{Berloff} and for its resemblance to the $\Tanh(x)$ function as shown in Fig.~\ref{fig:pade_approx}.   
It is important to keep in mind that if the $a_1$ is greater than $b_1$ the Pad\'e approximant loses its functional resemblance to $\Tanh(x)$ because the derivative,
\begin{align}\label{pade_deriv}
\small{
\frac{d}{dr} \sqrt{\frac{a_2 r^4+a_1 r^2}{a_2 r^4+b_1 r^2+1}} = \frac{a_2 r^5 (b_1-a_1)+a_1 r+2 a_2 r^3}{\left(a_2 r^4+b_1 r^2+1\right)^2 \sqrt{\frac{\left(a_1 r^2 + a_2 r^4\right)}{a_2 r^4+b_1 r^2+1}}},
}
\end{align}
\normalsize 
is negative as $r \to \infty$ causing the Pad\'e approximant to approach $1$ from above instead of from below as shown in Fig.~\ref{fig:bad_pade_approx}.  
To find a stationary initial condition, the Pad\'e trial solution (\ref{pade_form}) is inserted into (\ref{eomspin2}) where it is expand it in powers of $r$ until we can find the coefficients $a_1$, $a_2$, and $b_1$ that satisfy the equation of motion. Since the entire quadrupole is in one channel there is only one non-trivial equation we must satisfy.  For example, the equation for $m_f = 1$ is
\begin{align}
\nonumber
 r^3 \psi_1^3 \left(g_0 +  g_1\right) &- r E + \frac{\psi_1}{r} 
 \\
& - \frac{c_3 a_2^2 r^4  + c_2 a_1 a_2 r + c_1 a_1^2}{r \left(a_2 r^2+a_1\right){}^{3/2} \left(a_2 r^4+b_1 r^2+1\right){}^{5/2}} = 0,
\label{pade_long}
\end{align}
where, 
\begin{subequations}
\begin{align}
c_1 &= a_2^2 r^8-2 b_1 r^2+1 2  - a_2 r^4 \left(b_1 r^2+5\right)\\ 
c_2 &= a_2^2 r^8+b_1^2 r^4+b_1 r^2+3 -a_2 r^4 \left(b_1 r^2+8\right) \\ 
c_3 &= b_1^2 r^4+2 b_1 r^2+4-2 a_2 r^4 \left(b_1 r^2+4\right).
\end{align}
\end{subequations}
We can expand (\ref{pade_long}) in powers of $r$, and set the three lowest terms equal to zero to solve for $a_1$, $a_2$, and $b_1$.  Continuing the  the $m_f = 1$ dark vortex as an example the lowest order equations are
\begin{widetext}
\begin{align}
 &0 = \frac{\left(a_1 E -4 a_1 b_1+4 a_2\right)}{\sqrt{a_1}} \\ 
 &0 = \frac{\left(a_1^2 \left(16 a_2+b_1 \left(E -12 b_1\right)\right)-a_2 a_1 \left(E -8 b_1\right)+2 a_1^3 \left(g_0+g_1\right)+4 a_2^2\right)}{2 a_1^{3/2}} \\ \nonumber
&0 = \frac{1}{8 a_1^{5/2}}\Bigl(20 a_2^3 -a_1 a_2^2 \left(E -20 b_1\right)-2 a_1^2 a_2 \left(40 a_2+b_1 \left(E -30 b_1\right)\right)\\ \nonumber
& \hspace{1cm} -\left.a_1^3 \left(4 a_2 \left(-60 b_1+3 g_0+3 g_1+E \right)+b_1^2 \left(100 b_1-3 E \right)\right)\right) \\ 
& \hspace{1cm} + 12 a_1^4 b_1 \left(g_0+g_1\right)\Bigr)
\end{align}
\end{widetext}
\normalsize
which are the equations for the first, third, and fifth order in $r$ respectively.  When we solve these equations for $a_1$, $a_2$, and $b_1$ we get
\begin{subequations}
\begin{align}
a_2 &= a_1 \left(b_1 - \frac{E}{4}\right) \\ 
b_1 &= \frac{16 a_1 \left(g_0+g_1-3 E \right)+5 E ^2}{48 \left(E -4 a_1\right)} \\ 
a_1 &= \frac{11 E ^4}{76 E ^3-48 \left(g_0+g_1\right) E ^2 \pm  \sqrt{\gamma}}
,
\end{align}
\end{subequations}
where
\begin{align}
\gamma = E ^4 \left(361 E ^2-8 \left(g_0+g_1\right) \left(4 g_0+4 g_1+13 E \right)\right). \\ \nonumber
\end{align}
Since  there are only odd powers of $r$ in the expansion, there is no $r^6$ term.  This makes the Pad\'e approximant solution accurate to ${\cal O}(r^7)$. The coefficients  $a_1$, $a_2$, $b_1$ of the Pad\'e approximant for every single channel dark soliton solution can be found in Table~\ref{tab:Pade_sols}.
\begin{table}[!h!t!b!p]
\begin{center}
\begin{tabular}{|c|c|c|c|}
\hline
\multicolumn{4}{|c|}{Single channel Pad\'e approximant solutions} \\
\hline
\hspace{.1cm} $m_f$ \hspace{.1cm} & $a_1$ & $b_1$ & $a_2$ \\
\hline
& & &  \\ 
2
& $\frac{19 E ^3-12 \left(g_0+4 g_1\right) E ^2 \pm \gamma_2}{64 \left(g_0+4 g_1\right) \left(2 E -g_0-4 g_1 \right)}$ 
& $\frac{16 a_1 \left(g_0+4 g_1-3 E \right)+5 E ^2}{48 \left(E -4 a_1\right)}$
& $a_1 \left(b_1 - \frac{E}{4}\right)$\\ [.25cm]
\hline
& & &  \\ 
1
&$\frac{-11 E ^4}{76 E ^3-48 \left(g_0+g_1\right) E ^2 \pm 4\gamma_1}$
&$\frac{16 a_1 \left(g_0+g_1-3 E \right) + 5 E ^2}{48 \left(E -4 a_1\right)} $
&$a_1 \left(b_1 - \frac{E}{4}\right)$ \\[.25cm]
\hline
& & &  \\ 
0
&$\frac{5 \left(95 E^3 -12 \left(5 g_0+g_2\right) E ^2 \pm \gamma_0 \right)}{64 \left(5 g_0+g_2\right) \left(10 E -5 g_0-g_2 \right)}$
&$\frac{16 a_1 \left(5 g_0+g_2-15 E \right)+25 E ^2}{240 \left(E -4 a_1\right)}$
&$a_1 \left(b_1 - \frac{E}{4}\right)$\\[.25cm]
\hline
& & &  \\ 
1
&$\frac{-11 E ^4}{76 E ^3-48 \left(g_0+g_1\right) E ^2 \pm 4\gamma_1}$
&$\frac{16 a_1 \left(g_0+g_1-3 E \right) + 5 E ^2}{48 \left(E -4 a_1\right)} $
&$a_1 \left(b_1 - \frac{E}{4}\right)$ \\[.25cm]
\hline
& & &  \\ 
-2
& $\frac{19 E ^3-12 \left(g_0+4 g_1\right) E ^2 \pm \gamma_2}{64 \left(g_0+4 g_1\right) \left(2 E -g_0-4 g_1 \right)}$ 
& $\frac{16 a_1 \left(g_0+4 g_1-3 E \right)+5 E ^2}{48 \left(E -4 a_1\right)}$
& $a_1 \left(b_1 - \frac{E}{4}\right)$\\ [.25cm]
\hline
\end{tabular}
\end{center}
\caption{\label{tab:Pade_sols}\footnotesize  A table of the coefficients for the Pad\'e approximant that solves the spin-2 BEC equation to ${\cal O}(r^{7})$ for each $m_f$ level.  In these solutions  $\gamma_2 = E^2\sqrt{361 E ^2-8 \left(g_0+4 g_1\right) \left(4 g_0+16 g_1+13 E \right)}$, $\gamma_1 = E^2\sqrt{361 E ^2-8 \left(g_0+g_1\right) \left(4 g_0+4 g_1+13 E \right)}$, and~$\gamma_0 = E^2\sqrt{9025 E ^2-8 \left(5 g_0+g_2\right) \left(20 g_0+4 g_2+65 E \right)}$. }
\end{table}
\subsection{multichannel Pad\'e dark soliton}
To find two channel Pad\'e approximants it helps reduce the complexity if each channel is given the same Pad\'e approximant with the same unknown coefficients $a_1$, $a_2$, and $b_1$. For example our Pad\'e approximant for the 3 channel solution is given by 
\begin{align}
\Psi(r, \theta, t) 
=
\sqrt{\frac{a_2 r^4+a_1 r^2}{a_2 r^4+b_1 r^2+1}}e^{i \theta}e^{i E t} 
\begin{pmatrix}
1 \\
0 \\
1\\
0 \\
1 \\
\end{pmatrix}.
\end{align}
Following the procedure we outlined for the single channel Pad\'e approximant, we  end up with three sets of identical equations in the $m_f = 2, 0,$ and $-2$ channels.  The equations for the for the $r^1$, $r^3$ and $r^5$ terms of the expanded GPE are given by
\begin{subequations}
\label{three_pade}
\begin{align}
 a_1 E -4 a_1 b_1 + 4 a_2 &= 0 \\ 
 48 a_1 \left(5 E -20 b_1-5 g_0-g_2 \right) + 5 E  \left(48 b_1-5 E \right) &= 0 \\ 
\nonumber
 40 a_1 \mu ^2 \left(95 E -180 g_0-36 g_2 \right) + 275 E ^4 - \\ 
 \hspace{2cm} 768 a_1^2 \left(5 g_0+g_2\right) \left(-15 g_0-3 g_2+10 E \right) &= 0.
\end{align}    
\end{subequations}
Solving (\ref{three_pade}) for $a_1$, $a_2$ and $b_1$ gives 
\begin{subequations}
\begin{align}
a_2 &= a_1 \left(b_1 - \frac{E}{4}\right) \\ 
b_1 &=\frac{25 E ^2-48 a_1 \left(5 E-5 g_0-g_2 \right)}{240 \left(E -4 a_1\right)} \\ 
a_1 &= \frac{5 \left(95 E ^3 -36 \left(5 g_0+g_2\right) E ^2 \pm\gamma \right)}{192 \left(5 g_0+g_2\right) \left(10 E -15 g_0-3 g_2 \right)},
\end{align}
\end{subequations}
where,
\begin{equation}
\gamma = E^2\sqrt{9025 E ^2-24 \left(5 g_0+g_2\right) \left(60 g_0+12 g_2+65 E \right)}.
\end{equation}
In summary, we have found three different multichannel solutions for the Pad\'e dark soliton which are listed in Table \ref{tab:multi_Pade_sols}. 
\begin{table}[!h!t!b!p]
\begin{adjustbox}{width=\columnwidth,center}
\begin{tabular}{|c|c|c|c|}
\hline
\multicolumn{4}{|c|}{multichannel Pad\'e approximant solutions} \\
\hline
$m_f$'s & $a_1$ & $b_1$ & $a_2$ \\
\hline
& & &  \\ 
\small{1, -1}
& $\frac{5 \left(95 E ^3-24 \left(5 g_0+g_2\right) E ^2+\gamma_{\pm} \right)}{256 \left(5 g_0+g_2\right) \left(5 E -5 g_0-g_2 \right)}$ 
& $\frac{a_1 \left(160 g_0+32 g_2-240 E \right)+25 E ^2}{240 \left(E -4 a_1\right)}$
& $a_1 \left(b_1 - \frac{E}{4}\right)$\\ [.25cm]
\hline
& & &  \\ 
\small{2, -2}
& $\frac{5 \left(95 E ^3-24 \left(5 g_0+g_2\right) E ^2+\gamma_{\pm} \right)}{256 \left(5 g_0+g_2\right) \left(5 E -5 g_0-g_2 \right)}$ 
& $\frac{a_1 \left(160 g_0+32 g_2-240 E \right)+25 E ^2}{240 \left(E -4 a_1\right)}$
& $a_1 \left(b_1 - \frac{E}{4}\right)$\\ [.25cm]
\hline
& & &  \\ 
\small{2, 0, -2}
&$ \frac{5 \left(95 E ^3 -36 \left(5 g_0+g_2\right) E ^2 \pm\gamma_{3} \right)}{192 \left(5 g_0+g_2\right) \left(10 E -15 g_0-3 g_2 \right)}$
&$\frac{25 E ^2-48 a_1 \left(5 E-5 g_0-g_2 \right)}{240 \left(E -4 a_1\right)} $
&$a_1 \left(b_1 - \frac{E}{4}\right)$\\[.25cm]
\hline
\end{tabular}
\end{adjustbox}
\caption{\footnotesize  A table of the coefficients for the Pad\'e approximant that solves the spin-2 BEC equation to ${\cal O}(r^{7})$ for each $m_f$ level.  In these solutions  $\gamma_3 = E^2\sqrt{9025 E ^2-24 \left(5 g_0+g_2\right) \left(60 g_0+12 g_2+65 E \right)}$ and $\gamma_{\pm} = E^2 \sqrt{9025 E ^2-16 \left(5 g_0+g_2\right) \left(40 g_0+8 g_2+65 E \right)}$.   All three of these solutions represent the topological quantum vortex solitons with local quantum entanglement within the 5-dimensional Zeeman manifold of a spin-2 superfluid.}
\label{tab:multi_Pade_sols}
\end{table}  
We can classify the Pad\'e approximant solutions the same way we did the one dimensional solutions  where we determined which of $|\bm{F}_\parallel|$,  $|\bm{F}_\bot|$, or  $|\bm{A_{00}}|$, as defined in (\ref{cmap}), is the dominant term in the Gross-Pitaevskii equation. Table \ref{tab:Sol-Chars-Pade} shows a summary of those results.
\begin{table}[!h!t!b!p]
\begin{center}
\begin{tabular}{|c|c|c|c|}
\hline
\multicolumn{4}{|c|}{Solution Characteristics} \\
\hline
$m_f$'s  & $|\bm{F}_\parallel|$ & $|\bm{F}_\bot|$ & $|\bm{A_{00}}|$\\
\hline 
& & & \\ 
2 & $\frac{2 \left(a_1 r^2+a_2 r^4\right)}{a_2 r^4+b_1 r^2+1}$ & 0 & 0  \\[.2cm]
1 &$\frac{a_1 r^2+a_2 r^4}{a_2 r^4+b_1 r^2+1}$& 0 & 0 \\[.2cm]
0 & 0 & 0 &$\frac{a_1 r^2+a_2 r^4}{\sqrt{5} \left(a_2 r^4+a_1 r^2+1\right)}$  \\[.2cm]
-1 &$\frac{a_1 r^2+a_2 r^4}{a_2 r^4+b_1 r^2+1}$& 0 & 0  \\[.2cm]
-2 & $\frac{2 \left(a_1 r^2+a_2 r^4\right)}{a_2 r^4+b_1 r^2+1}$ &  0 &0 \\[.2cm]
1, -1 & 0 & 0 & $\frac{2 \left(a_1 r^2+a_2 r^4\right)}{\sqrt{5} \left(a_2 r^4+b_1 r^2+1\right)}$ \\[.2cm]
2, -2 & 0 & 0 & $ \frac{2 \left(a_1 r^2+a_2 r^4\right)}{\sqrt{5} \left(a_2 r^4+b_1 r^2+1\right)} $\\[.2cm]
2,0,-2  &  0 & 0 & $\frac{3 \left(a_1 r^2+a_2 r^4\right)}{\sqrt{5} \left(a_2 r^4+b_1 r^2+1\right)}$ \\[.2cm]
\hline
\end{tabular}
\end{center}
\caption{\footnotesize  The characteristics of the Pad\'e approximant solutions as a function of $r$. $a_1$, $a_2$, and $b_1$ are the coefficients for a Pad\'e approximant of form (\ref{pade_form}) and can be found in Tables~\ref{tab:Pade_sols} and \ref{tab:multi_Pade_sols}.}
\label{tab:Sol-Chars-Pade}
\end{table}

\section{Quantum simulations in 2D}
\label{Section_Quantum_simulations_in_2D}

Once again we perform calibration experiments as well as scattering (soliton-soliton interaction) experiments.  In two dimensions the calibration experiments cannot just be a single dark soliton, since if we naively placed a single dark vortex in the center of the lattice the repeating boundary conditions would cause a discontinuity in the phase at the boundary.  That is to say the quantum fluid is moving in opposite directions at opposing edges. This is not a physically interesting state, and it is an unacceptable initial condition for the quantum lattice gas algorithm.   
\paragraph*{}
A way to circumvent this nonperiodicity issue is to place four vortices in the lattice located equidistant from each other in a quadrupole formation with two clockwise rotating vortices along one diagonal and two counter clockwise rotating vortices along the other diagonal.  This restores repeating boundary conditions.
\subsection{Calibration Experiments in 2D}
\subsubsection{Stationary Quadrupole}
\begin{figure}[!h!t!b!p]
\begin{center}
\includegraphics[width= 1\linewidth]{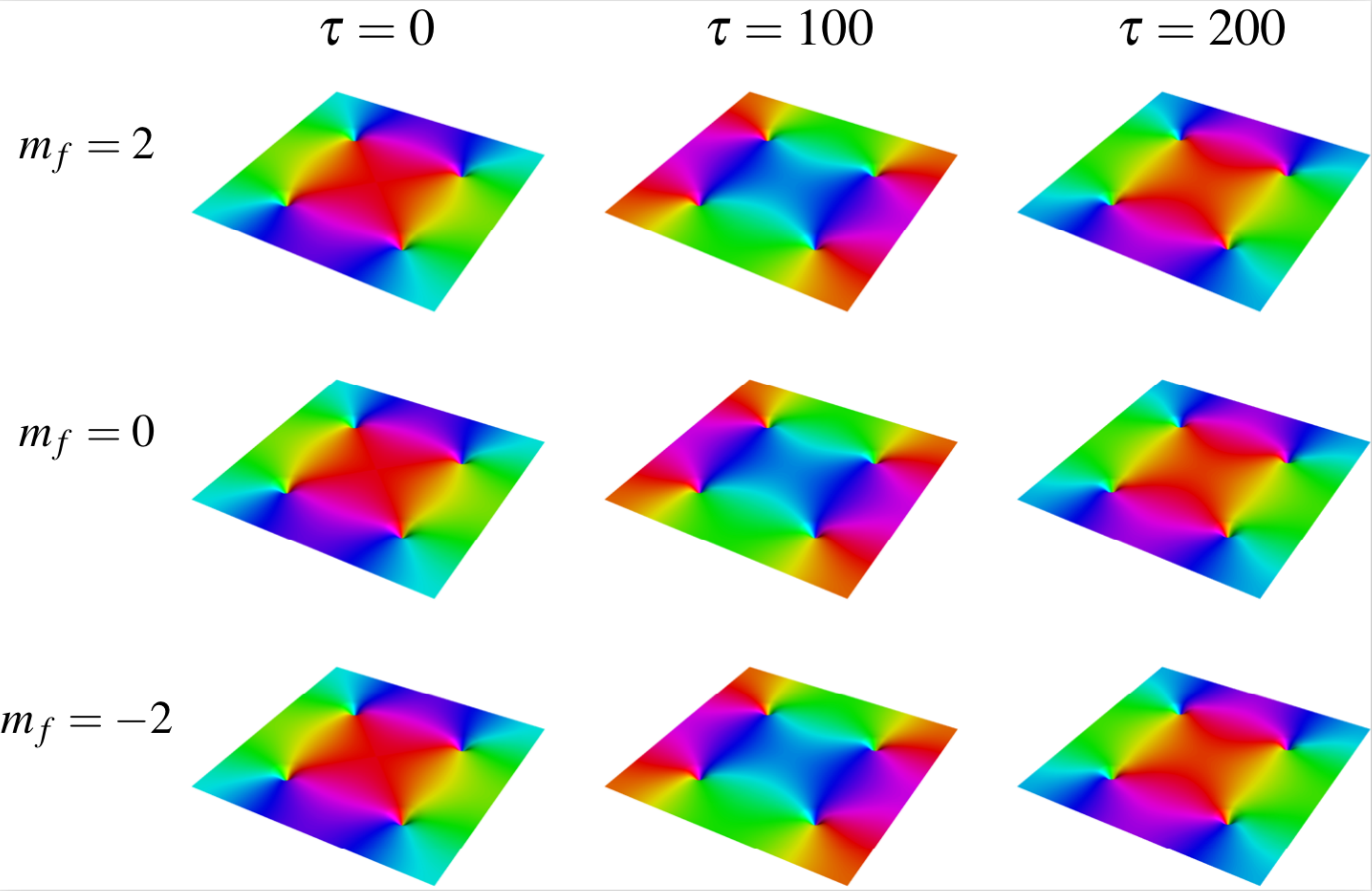} 
\end{center}
\caption{\footnotesize  A stationary three channel quadrupole of Pad\'e approximant vortices in the $m_f = 2$, $m_f = 0$, and $m_f = -2$  channels of a spin-2 BEC spinor superfluid.  The color on the plot represents the phase of the quantum field and the height is the density of the field.  Since the velocity a quantum fluid with uniform density is just the gradient of the phase, the velocity the fluid in the bulk (away from the vortex center) is simply the color gradient.  Shown here is the initial condition, the vortex after one half rotation and then the vortex again after a full rotation. The period of a full revolution is approximately 200 time steps.}%
\label{fig:stationary_three_channel}
\end{figure}

The first calibration experiment is a stationary quadrupole where the four vortices rotate but do not have any total momentum. If we modify the algorithm as is outlined in Section~\ref{trotterization}. We can achieve stable quadrupole on a lattice as small as 512 x 512 sites. Keep in mind that since we assumed that $\psi_{m_f}$ we should pick $E$, $g_0$, $g_1$, and $g_2$ such that $a_1$, $a_2$ and $b_1$ are all positive additionally $b_1$ must be greater than $a_1$ to maintain the Pad\'e approximant's resemblance to the dark soliton $\Tanh(x)$ form.  A visualization of a single channel these initial conditions is given in Fig.~\ref{fig:pade_init_stationary}.  
\begin{figure}[!h!t!b!p]
\label{fig:pade_init_stationary}
\begin{center}
\includegraphics[width= 1\linewidth]{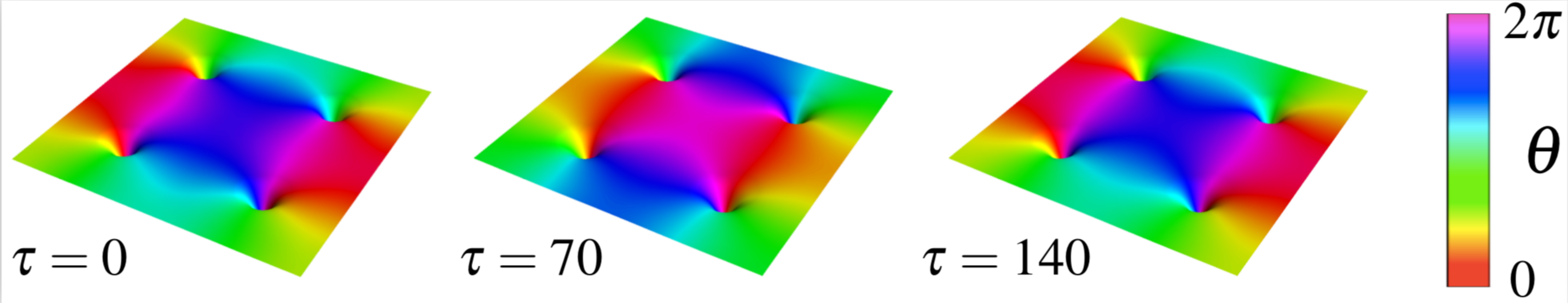}
\end{center}
\caption{\footnotesize  A stationary quadrupole of Pad\'e approximant vortices in the $m_f = 1$ channel of a spin-2 BEC spinor superfluid.  The color on the plot represents the phase of the quantum field and the height is the density of the field.  Since the velocity a quantum fluid with uniform density is just the gradient of the phase, the velocity the fluid in the bulk (away from the vortex center) is simply the color gradient.  Shown here is the initial condition, the vortex after one half rotation and then the vortex again after a full rotation. The period of a full revolution is approximately 140 time steps.}\label{fig:pade_init_stationary}
\end{figure}
The multichannel Pad\'e approximant vortices are also stable.  A visualization of the three channel vortex is shown in  Fig.~\ref{fig:stationary_three_channel}.
All eight Pad\'e vortices are stable and can be run indefinitely using the quantum lattice gas algorithm.  This is an important check of both the quantum lattice gas algorithm and the Pad\'e aprroximant vortex solutions. This is paramount because once we start colliding vortices we will have no analytic solution to test against so we must make absolutely sure the vortices are behaving as predicted while we have the luxury of an analytical solution.

\subsubsection{Kicked Quadrupole}
\begin{figure}[!h!t!b!p]
\begin{center}
\includegraphics[width= 1\linewidth]{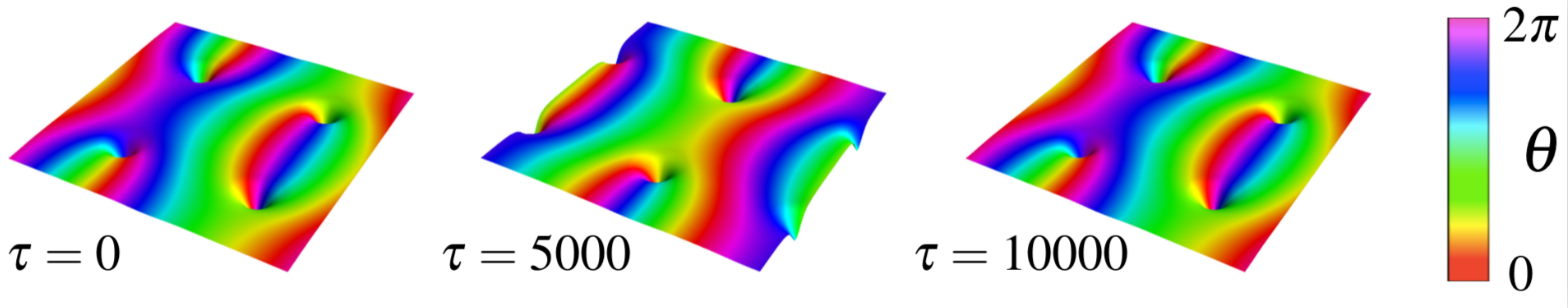}
\end{center}
\caption{\label{fig_pade_init_kicked}
\footnotesize  A kicked Pad\'e approximant vortices in the $m_f = 1$ channel  of a spin-2 BEC spinor superfluid.  Once again the color on the plot represents the phase of the quantum field and the height is the density of the field and the velocity in the bulk is the color gradient.  Note how the fluid is moving in a circle around the dark vortex core as well as moving along the $x$ direction. One vortex moves across half the lattice, to the initial location of a different vortex in approximately 5000 time steps.}
\end{figure}

Another important check we need to make before we collide vortices is to ensure that a kicked quadrupole is stable.  A kicked quadrupole has the same initial conditions as the stationary quadrupole multiplied by $e^{i({2 \pi n x}/{L})}$ or $e^{i({2 \pi n y}/{L})}$, where $L$ is the length of the lattice along the $x$ or $y$ directions respectively.  This both maintains the smoothness at the boundary and gives a momentum to the entire field in the $x$ direction.  Fig.~\ref{fig_pade_init_kicked} shows the initial conditions of a quadrupole kicked in the $x$ direction with $n=2$ so the phase repeats twice across the lattice and is the second slowest allowed speed. 

\begin{figure}[!h!t!b!p]
\begin{center}
\includegraphics[width= 1\linewidth]{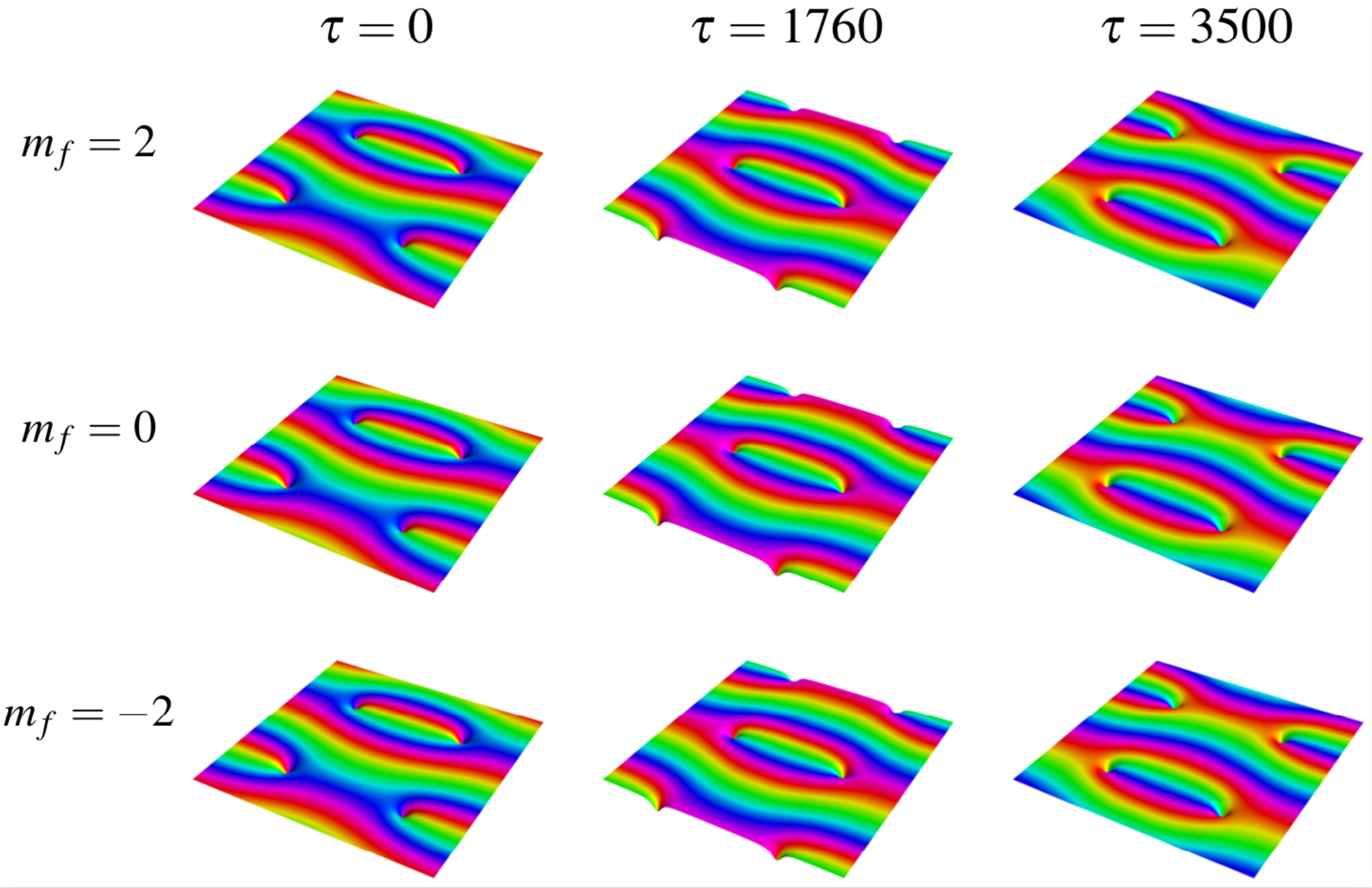} 
\end{center}
\caption{\footnotesize  A kicked three channel quadrupole in the $m_f = 2$,  $m_f = 0$, and $m_f = -2$ channels  of a spin-2 BEC spinor superfluid.  Once again the color on the plot represents the phase of the quantum field and the height is the density of the field and the velocity in the bulk is the color gradient.  Note how the fluid is moving in a circle around the dark vortex core as well as moving along the $y$ direction. One vortex moves across half the lattice, to the initial location of a different vortex in approximately 3500 time steps.}
\label{fig:kicked_three_channel}
\end{figure}

The three channel Pad\'e approximant vortex is also stable as shown in Fig.~\ref{fig:kicked_three_channel}.  This time we kicked the vortex in the $y$ direction with $n = 3$.  Notice how the gradient of the phase in the bulk has changed direction and the phase repeats an additional time.   
All eight of the Pad\'e approximant vortices are stable and will traverse the lattice indefinitely in both the $x$ and $y$ directions.  Now that we know our solutions are indeed stable and can be given momenta in any direction we are  ready to scatter (collide) two sets of quadrupoles.  However, in order to better understand the nature of the interactions it is crucial to look at the dynamics of the vortices, and to understand the dynamics of the vortices we must first be able to determine where the vortices are at any given time.
\subsection{Calculating quantum vorticity on a lattice}
Classically the vorticity $\bm{\omega}$ is the curl of the velocity field 
\begin{equation}\label{vorticity}
\bm{\omega}  = \nabla \cross \bm{v}. 
\end{equation}   
For a spinor superfluid with multiplet field 
\begin{align}
\psi_m = |\psi_m| e^{i \arg \psi_m}
,
\end{align}
 a ``classical" velocity field $\bm{v}_m$ can be calculated for the $m^\text{th}$ hyperfine level of the Zeeman manifold using the momentum operator $\hat{\bm{p}} = -i\hbar \nabla$ as
\begin{subequations}
\begin{align}
   \bm{v}_m
    \psi_m
   &\equiv
  \frac{ \nabla S_m}{ m}
    \psi_m
\\
&=   
  \frac{\hbar \nabla \arg \psi_m}{m}
      \psi_m
   ,
\end{align}
\end{subequations}
where the phase action is $S_m = \hbar \arg \psi_m$. Hence, the vorticity field of the  $m^\text{th}$ hyperfine level of the Zeeman manifold is determined by $\bm{\omega}_m  = \nabla \cross \bm{v}_m$.

Since the quantum lattice gas algorithm is run on a 2 dimensional lattice we do not need a vector vorticity, since the only nonzero vorticity will be pointing directly into or out of the lattice.  Additionally, we will be using a symmetric discretized version of the derivative 
\begin{equation}\label{discrete_deriv}
\frac{d}{dx} f(x) = \frac{f(x + 1) - f(x-1)}{2}. 
\end{equation}
This results in a scalar vorticity for the $m^\text{th}$ hyperfine level
\begin{equation}\label{pointwise_vorticity}
\omega(x, y) = \frac{v_x(x + 1, y) +  v_y(x, y - 1) - v_y(x-1, y) - v_x(x, y + 1)}{2},
\end{equation}     
where $v_x$ and $v_y$ are the usual given by a discretized version of the usual quantum fluid velocity given in  \cite{barenghi:2016}
\begin{subequations}
\begin{align}\label{velocity}
v_x(x, y) &= \frac{S (x+1, y) - S (x-1, y)}{2} 
\\ 
v_y(x, y) &= \frac{S (x, y+1) - S (x, y-1)}{2},
\end{align}
\end{subequations}
and $S(x,y)$ is the phase of the probability amplitude $\psi$ at the point $(x, y)$.  

\subsection{Quantum vortex soliton-soliton collisions in 2D}
\subsubsection{Dark soliton vortex collision}
\begin{figure}[!h!t!b!p]
\begin{center}
\includegraphics[width= 1\linewidth]{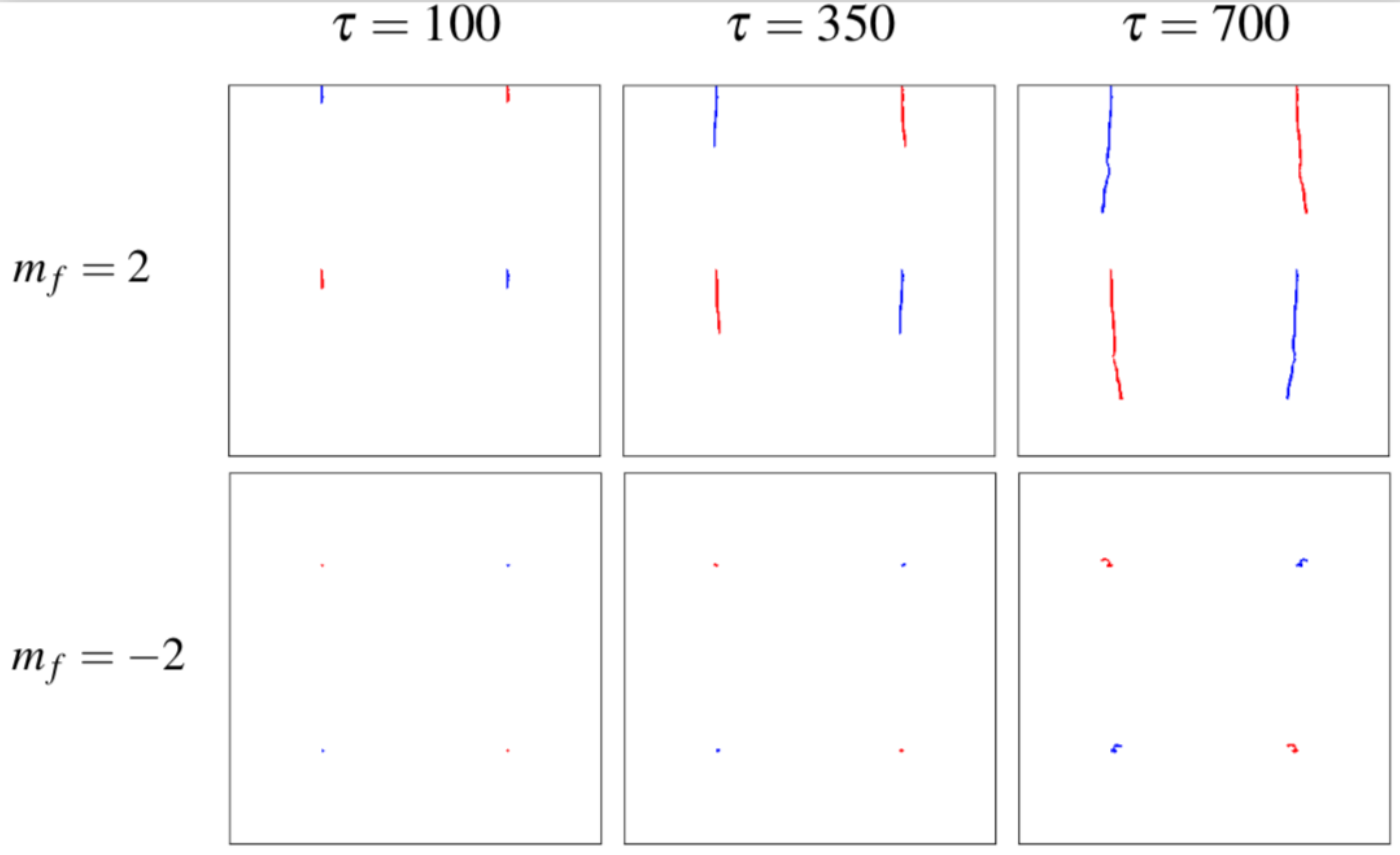} 
\end{center}
\caption{\footnotesize The motion of the vortices in the dark soliton collision between solitons in the $m_f = 2$ and $m_f = -2$ channels. The red tracks show quantum vortices with a positive winding number and the blue tracks show quantum vortices with negative winding number.}%
\label{fig:Quadrupole Collision Vortex Tracks}
\end{figure}

In the quadrupole collision experiments we set up one stationary quadrupole in one $m_f$ level and another kicked quadrupole field in a different hyperfine level offset in the $y$ direction by $L/4$.  Now that we have two different quadrupoles we need to be extra careful about choosing $\mu$, $g_0$, $g_1$, and $g_2$ since we need to have both quadrupoles meet the criteria $a_1$, $a_2$, $b_1$ > 0 and $b_1$ > $a_1$.  All the single channel Pad\'e apprroximants meet these criteria with $\mu = 1$, $g_0 = 1$  $g_1 = .1$ $g_2 = 1$.  As the vortices interact they can transfer momenta, create and annihilate vortices pairwise, and even excite new hyperfine levels. The first example we have is a simple scattering of a vortex in the $m_f = 2$ channel off of a stationary vortex in the  $m_f = -2$ channel.  In this example we see only momentum transfer as the $m_f = -2$ vortices get swirled around as the $m_f = 2$ vortices pass through.  In turn the trajectories of the $m_f = -2$ vortices are altered. %
\begin{figure}[!h!t!b!p]
\begin{center}
\includegraphics[width= 1\linewidth]{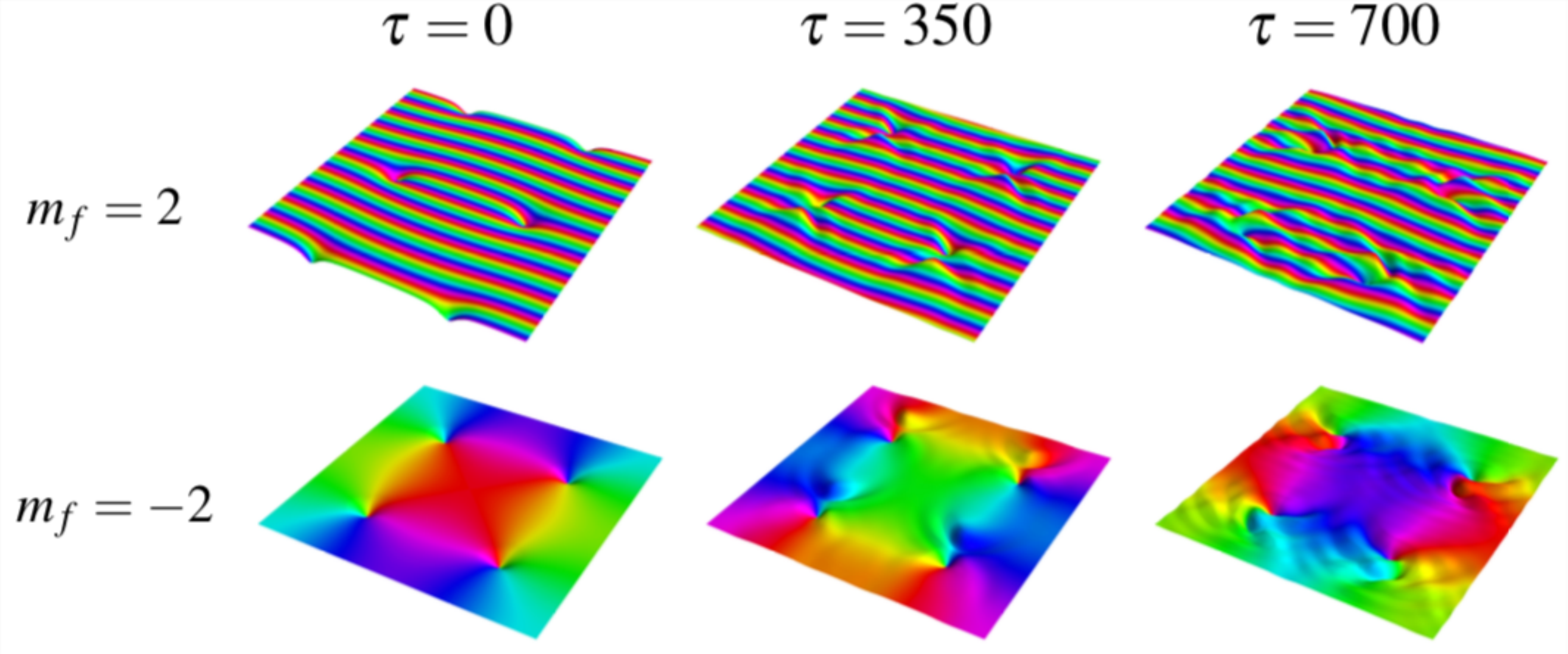} 
\end{center}
\caption{\footnotesize  A two dimensional dark soliton dark soliton collision between solitons in the $m_f = 2$ and $m_f = -2$ channels. The time of the collision is at approximately 500 time steps. The color of the image represents the phase with red corresponding to $0$ and purple corresponding to $2 \pi$ and the height of the image represents the density of the quantum fluid. }%
\label{fig:Quadrupole Collision}
\end{figure} 
 Fig.~\ref{fig:Quadrupole Collision} shows the collision between two sets of quadrupoles in the $m_f = 2$ and $m_f = -2$ channels.  The gradient of the phase means that the quadrupole in the $m_f = 2$ channel is headed towards the quadrupole in the $m_f = -2$ channel.  At $\tau = 350$ the dark solitons have just started to interact and you can see some ripples in the background field in each channel.  By $\tau = 700$ the dark solitons have passed by each other and the ripples in the field are more prevalent. The motion of the vortices is captured in Fig.~\ref{fig:Quadrupole Collision Vortex Tracks}, where the red tracks show quantum vortices with a positive winding number and the blue tracks show quantum vortices with negative winding number. 
 In the reference frame of the $m_f = 2$ dark soliton we see that a vortex with a positive winding number incident upon a vortex with a negative winding number causes a deflection to the left while a vortex with a negative winding number incident upon a vortex with a positive winding number causes a deflection to the right. 

\paragraph*{}
Next we look at a collision between an $m_f = 2$ and $m_f = 1$ dark solitons.  This collision features momenta transfer, vortex pair creation and annihilation, and excites all five $m_f$ levels.  Other than the $m_f$ channels of the dark solitons the simulation is set up in the same way as before. 
\begin{figure}[!h!t!b!p]
\begin{center}
\includegraphics[width= 1\linewidth]{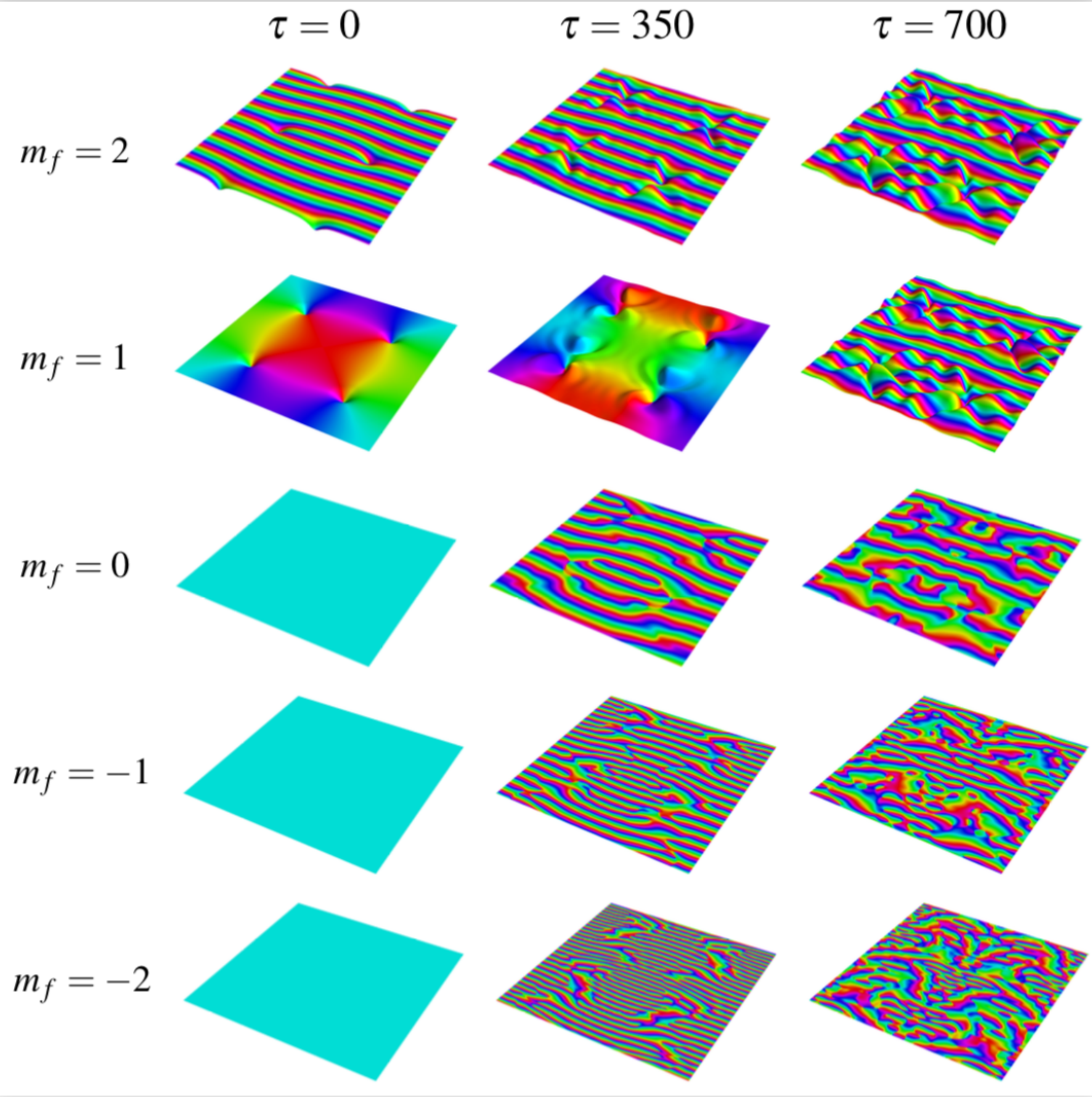} 
\end{center}
\caption{\label{fig:Quadrupole Collision complicated}\footnotesize  A two dimensional dark soliton collision between solitons in the $m_f = 2$ and $m_f = 1$ channels. The time of the collision is at approximately 500 time steps. The color of the image represents the phase with red corresponding to $0$ and purple corresponding to $2 \pi$ and the height of the image represents the density of the quantum fluid. }
\end{figure}
Fig.~\ref{fig:Quadrupole Collision complicated} shows production of complex quantum vortices via dark soliton-soliton scattering. A quadrupole kicked in the $y$ direction in the $m_f = 2$ channel scatters off of an offset but stationary quadrupole in the  $m_f = 1$ channel.  At early times the presence of a dark vortex in the $m_f = 1$ channel creates regions of high density in the $m_f = 2$ channel and vice-versa.  Meanwhile a moving low density field is imprinted in the originally empty  $m_f = 0$,  $m_f = -1$, and $m_f = -2$ channels.  The imprinted phase in the empty channels forms the same pattern only with a larger phase gradient in the $m_f = -2$ channel and a smaller phase gradient in the $m_f = 0$ channel.  Vortices in the low density channels vortices can be identified by a color pinwheel where the phase makes at least one full rotation about a point.  It is unclear, however, whether the vortices in the low density channels of Fig.~\ref{fig:Quadrupole Collision complicated} are bright or dark solitons.  Fig.~\ref{fig:complicated_collision_log} uses a logarithmic scale to show that the centers of these vortices have zero density and are therefore dark solitons.
\begin{figure}[!h!t!b!p]
\begin{center}
\includegraphics[width= 1\linewidth]{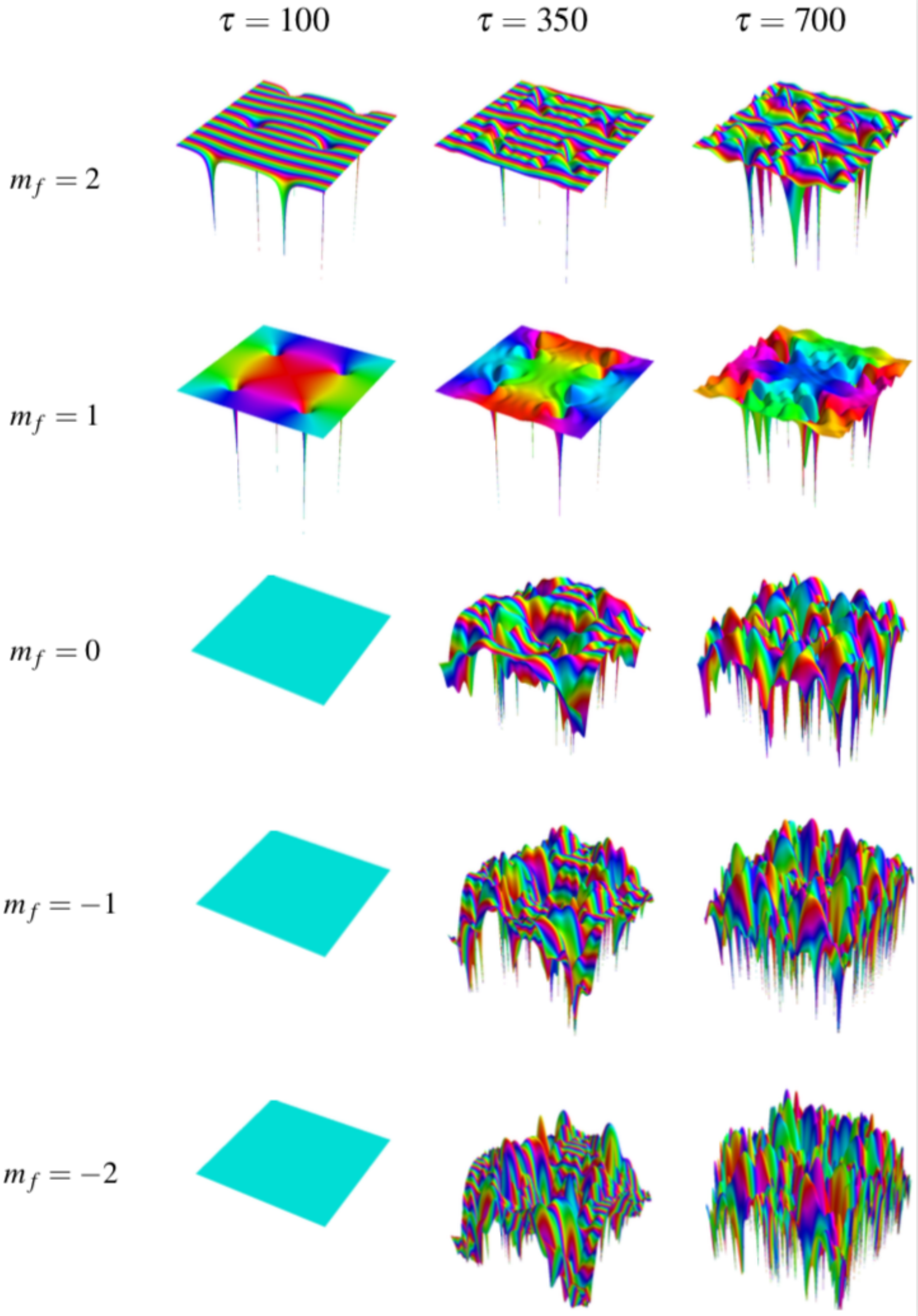} 
\end{center}
\caption{\footnotesize The exact same experiment as Fig.~\ref{fig:Quadrupole Collision complicated} only shown with a logarithmic scale of the density.  This highlights the fact that the vortex centers have zero density even in the low density fields which classifies the created solitons as dark soltions.}%
\label{fig:complicated_collision_log}
\end{figure} 
The dynamical behavior of these spin-2 dark soliton vortices is truly fascinating. The vortices that are created in the originally empty channels occur at the eight locations of the original vortices in the $m_f = 2$ and $m_f = 1$ channels.  The vortices in the  $m_f = 0$,  $m_f = -1$, and $m_f = -2$  channels with start with winding numbers 2, 3, and 4 respectively. By $\tau = 700$  the collisions between the $m_f = 2$ and $m_f = 1$ channels have each created a a second set of vortices in each channel while the $m_f = 0$,  $m_f = -1$, and $m_f = -2$ channels have created and annihilated many vortices. As demonstrated in  Fig.~\ref{fig:Quadrupole Collision Vortex Tracks Momentum},  every time a positive winding number vortex is created a corresponding negative winding number vortex is created thus conserving the total vortex winding number of the system.   
\begin{figure}[!h!t!b!p]
\begin{center}
\includegraphics[width= 1\linewidth]{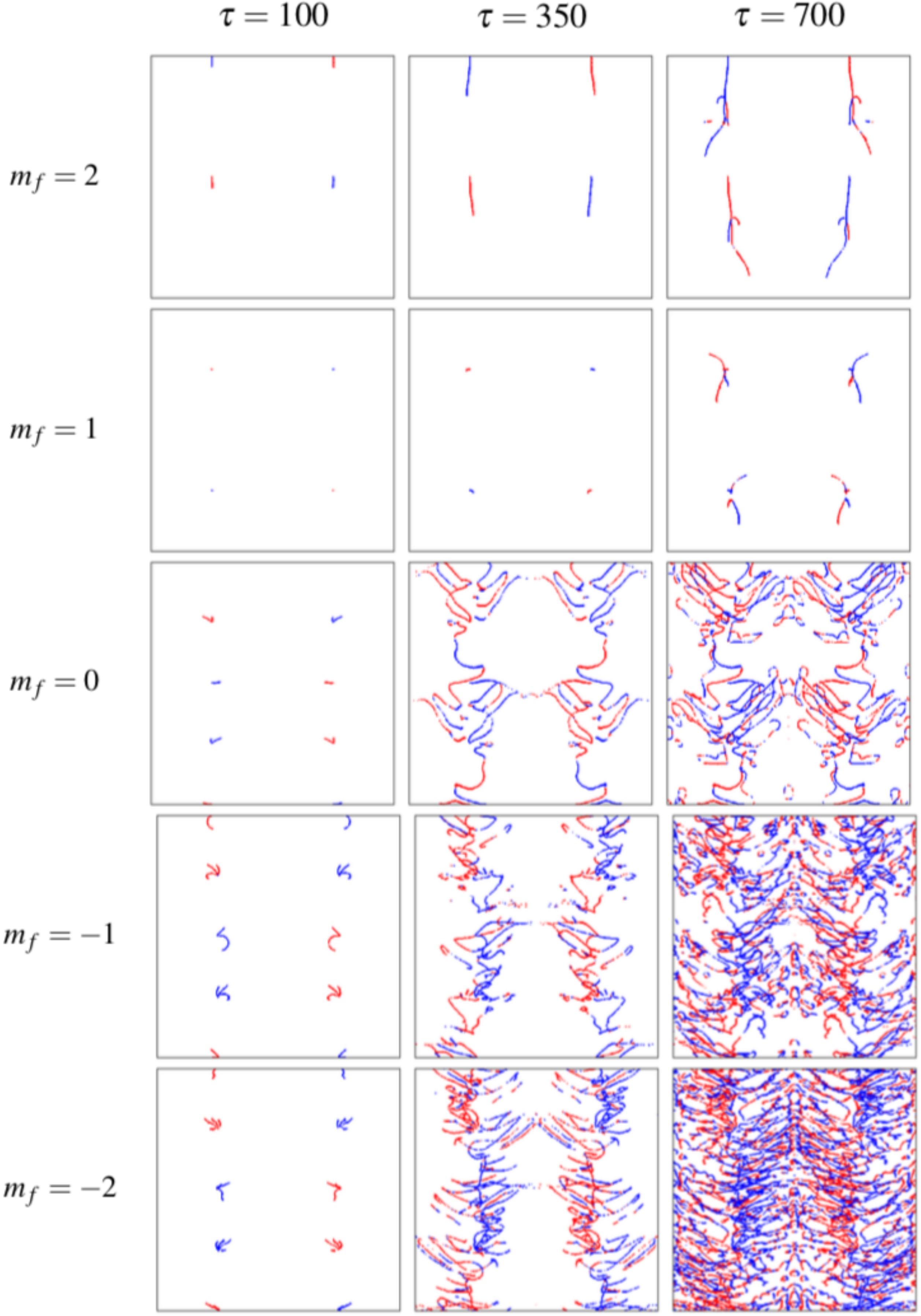} 
\end{center}
\caption{\footnotesize Production of complex quantum vortices via dark soliton-soliton scattering.  Initially there exists nonzero vorticity only in the $m_f=2$ and $m_f=1$ hyperfine levels, yet vorticity in all the hyperfine levels is rapidly generated indicating the emergence of complex quantum vortices.  The motion of the vortex centers is shown for the quantum vortices in experiment given in Fig.~\ref{fig:Quadrupole Collision complicated}. The red tracks are vortices with a positive winding number and the blue tracks are vortices with negative winding number.}%
\label{fig:Quadrupole Collision Vortex Tracks Momentum}
\end{figure} 
The excitation of vortices across the whole hyperfine manifold is unique to spinor BECs and is not and cannot exist in scalar BECs.  Furthermore, if you look closely at the vorticity patterns in the $m_f = 0$,  $m_f = -1$, and $m_f = -2$ you can see several loops of vorticity that have multiple segments of positive and negative vorticity. We have not seen this phenomena in scalar BECs. Fig.~\ref{fig:zoom_in} shows a magnified version of the $m_f = 0$, $\tau = 700$ portion of Fig.~\ref{fig:Quadrupole Collision Vortex Tracks Momentum} which illustrates this feature.  
\begin{figure}[!h!t!b!p]
\includegraphics[width= 0.8\linewidth]{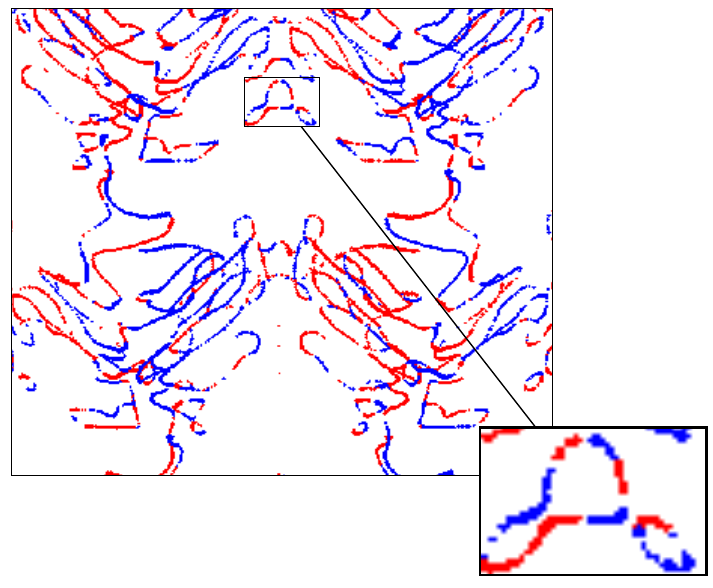} 
\caption{\footnotesize  A magnified version of the $m_f = 0$, $\tau = 700$ portion of Fig.~\ref{fig:Quadrupole Collision Vortex Tracks Momentum}.  The zoomed in portion shows two loops one positive and negative winding number vortex creation and annihilation, and another with three positive vorticity line segments and three negative vorticity line segments.}%
\label{fig:zoom_in}
\end{figure} 
\section{Quantum simulation implementation}
\label{Section_Quantum_simulation_implementation}

In order for the quantum lattice gas algorithm to run properly there are some critical details that need to be implemented to ensure a faithful reproduction of the physics.  It is also worth mentioning that the algorithm is implemented on general-purpose graphical processing units (GPUs) with error correcting code (ECC) memory to maximize the site update rate while maintaining double precision accuracy.   
\subsection{Scaling}
The algorithm is sensitive to rapid changes in the quantum field between adjacent lattice sites.  It is therefore critical that we are able to scale the spatial dimension so that we can ensure that the variation between lattice sites is always sufficiently small.  So, the question is if we let $x \to ax$ what other terms in the spinor Gross-Pitiaevskii (GP) equation do we need to rescale to recover an identical set of governing equations. Looking at  (\ref{eomspin2}) we find that there are three different types of terms: (1) terms proportional to $\nabla^2 \psi(x, t)$, (2) terms proportional to  $\partial_t \psi(x,t)$, and (3) terms proportional to $g_i$.  Thus we must introduce three scalings, $x \to ax$, $t \to \tau t$ and $g_i \to \gamma g_i$.  Inserting these scaled variables into  (\ref{eomspin2}) we get
\begin{align}\label{eomscaling}
\nonumber
&i \partial_t \left(\psi_m(ax, \tau t)\right) = -{\partial_{x}^2}\left( \psi_m(ax, \tau t)\right)  \\ \nonumber
&\hspace{2.2cm} + \gamma g_0 \left |\psi(ax, \tau t) \right |^2 \psi_m(ax, \tau t) \\ \nonumber
& \hspace{2.2cm}+ \gamma g_1 \sum_{m' = -2}^{m'=2}\bm{F} \cdot \bm{f}_{mm'}\psi_m(ax, \tau t) \\ 
&\hspace{2.2cm}+ \gamma g_2 \left|{A}_{00}\right|^2\psi^*_{-m}(ax, \tau t). 
\end{align}
Using the chain rule gives
\begin{align}\label{eomscaling}
\nonumber
&i \tau \partial_{\tau t} \left(\psi_m(ax, \tau t)\right) = -a^2 {\partial_{a x}^2} \psi_m(ax, \tau t)  \\ \nonumber
&\hspace{2.5cm} + \gamma g_0 \left |\psi(ax, \tau t) \right |^2 \psi_m(ax, \tau t) \\ \nonumber
& \hspace{2.5cm}+ \gamma g_1 \sum_{m' = -2}^{m'=2}\bm{F} \cdot \bm{f}_{mm'}\psi_m(ax, \tau t) \\ 
&\hspace{2.5cm}+ \gamma g_2 \left|{A}_{00}\right|^2\psi^*_{-m}(ax, \tau t),
\end{align}
which upon inspection is equivalent to  (\ref{eomspin2}) if $\tau = \gamma = a^2$.  It is worth noting that there are at least two ways to implement the $\gamma$ scaling.  One has the choice between $g_i \to \gamma g_i$ and $\psi_m~\to~\sqrt{\gamma} \psi_m$--- each way gives the same result of $\gamma = a^2$. In our specific implementation we used $\psi_m~\to~\sqrt{\gamma} \psi_m$. 
\subsection{Operator interleaving}\label{trotterization}
\subsubsection{Theoretical improvement}
A possible way to improve the quantum lattice gas algorithm's accuracy is to interleave the self-interaction operator with the stream and collide operators.  Let us define 
\begin{subequations}
\begin{align}
\mathbb{I}'_1  &=  \hat{S}_{1}\hat{C}\hat{S}^T_{1}\hat{C} \\ 
\mathbb{I}'_2 &=  \hat{S}_{2}\hat{C}\hat{S}^T_{2}\hat{C},
\end{align}
\end{subequations}
where the $\hat{S}$ and $\hat{C}$ operators are the stream and collide operators described in \cite{yepez:2002}.  We have given names like $\mathbb{I}'_i$
since they are approximately the identity operator when acting on the bosonic state $\psi$.  In fact, if you simply switched the order of  the stream and collide operators in $\mathbb{I}'_1$ or $\mathbb{I}'_2$ such that there was no interleaving of streams and collides one would exactly recover the identity operator.  The quantum lattice gas algorithm we used to simulate the spin-2 GP equation is given by
\begin{align}
\psi(t + \tau) = (\mathbb{I}'_1\mathbb{I}'_1
\mathbb{I}'_2\mathbb{I}'_2
\hat{U})\psi(t),
\end{align} 
where $\hat{U}$ is the self-interaction operator detailed in \cite{yepez:2016} and $\tau$ is a single time step. This would make $\mathbb{I}'_1\mathbb{I}'_1\mathbb{I}'_2\mathbb{I}'_2$ the kinetic energy operator, and indeed if you expand it out you get
\begin{align}
(\mathbb{I}'_1\mathbb{I}'_1\mathbb{I}'_2\mathbb{I}'_2)\psi = \psi - i \hat{p}^2 + {\cal O}(\hat{p}^4),
\end{align}
which is exactly the kinetic energy operator until fourth order.  While the time evolution of the spin-2 BEC is generally given by
\begin{equation}
\psi(t +\tau) = e^{-i(\hat{p}^2 + \hat{U})\tau}\psi(t),
\end{equation}  
the quantum lattice gas algorithm simply applies $\hat{U}$ after applying the kinetic energy operator.   This can introduce error since 
\begin{equation}
e^{A+B} \neq e^A e^B,
\end{equation}
unless $A$ and $B$ commute.  We can reduce this error by applying $\hat{U}^{1/4}$ after each $\mathbb{I}'$ operator giving
\begin{align}
\psi(t +\tau)  = (\mathbb{I}'_1\hat{U}^{1/4}\mathbb{I}'_1\hat{U}^{1/4}
\mathbb{I}'_2\hat{U}^{1/4}\mathbb{I}'_2\hat{U}^{1/4}) \psi(t).
\end{align}  
This helps because any operator commutes with $\mathbb{I}$ so combining $\hat{U}^{1/4}$ with $\mathbb{I}'_i$ will give a good approximation for one quarter of our desired time evolution,   that is to say 
\begin{equation}
\mathbb{I}'_i \hat{U}^{1/4} \approx e^{\frac{p^2 + U}{4}}
.
\end{equation}
In addition to the near commutativity of $\hat{U}^{1/4}$ and  $\mathbb{I}'_i$ we get some additional accuracy from the resemblance of our evolution algorithm to the $n=4$ version Lie product formula \cite{lie_engel_1970}
\begin{equation}
e^{(A+B)}=\lim_{n\to\infty} \left(e^{\frac{A}{n}}e^{\frac{B}{n}}\right)^n.
\end{equation} 
In two and three dimensions the kinetic part of the quantum lattice gas algorithm has more $\mathbb{I}'_i$ operators so we can use the Lie product formula with $n=8$, and $n=12$  respectively.  Trying to interleave $U$ in the middle of a $\mathbb{I}'_i$ operator will result in the break down the quantum lattice gas algorithm.      

\subsubsection{Practical improvement}

Using this modified quantum lattice gas algorithm is paramount in creating stable simulations with a non-zero background.  For example, the dark soliton stationary states would be stable for around 40,000 time steps without the operator splitting, but can run stably for more than 2 million time steps with the split operator algorithm. This gives us a much wider window to explore the dynamics of the spin-2 BEC.

Another advantage of interleaving the quantum lattice gas algorithm is that it can speed up simulation time.  For example in two dimensions using the interleaved algorithm allows us to run a stable Pad\'e approximant representation of a vortex quadrupole on a lattice as small as 512 x 512 sites. The non-interleaved algorithm required 4048 x 4048 site lattice to run the same simulation.  Which is 64 times fewer sites and allows the simulation to run 64 times faster which is a significant speed up. 

\begin{figure}[!h!t!b!p]
\begin{center}
\hspace{-.1cm}\includegraphics[width=.8\linewidth]{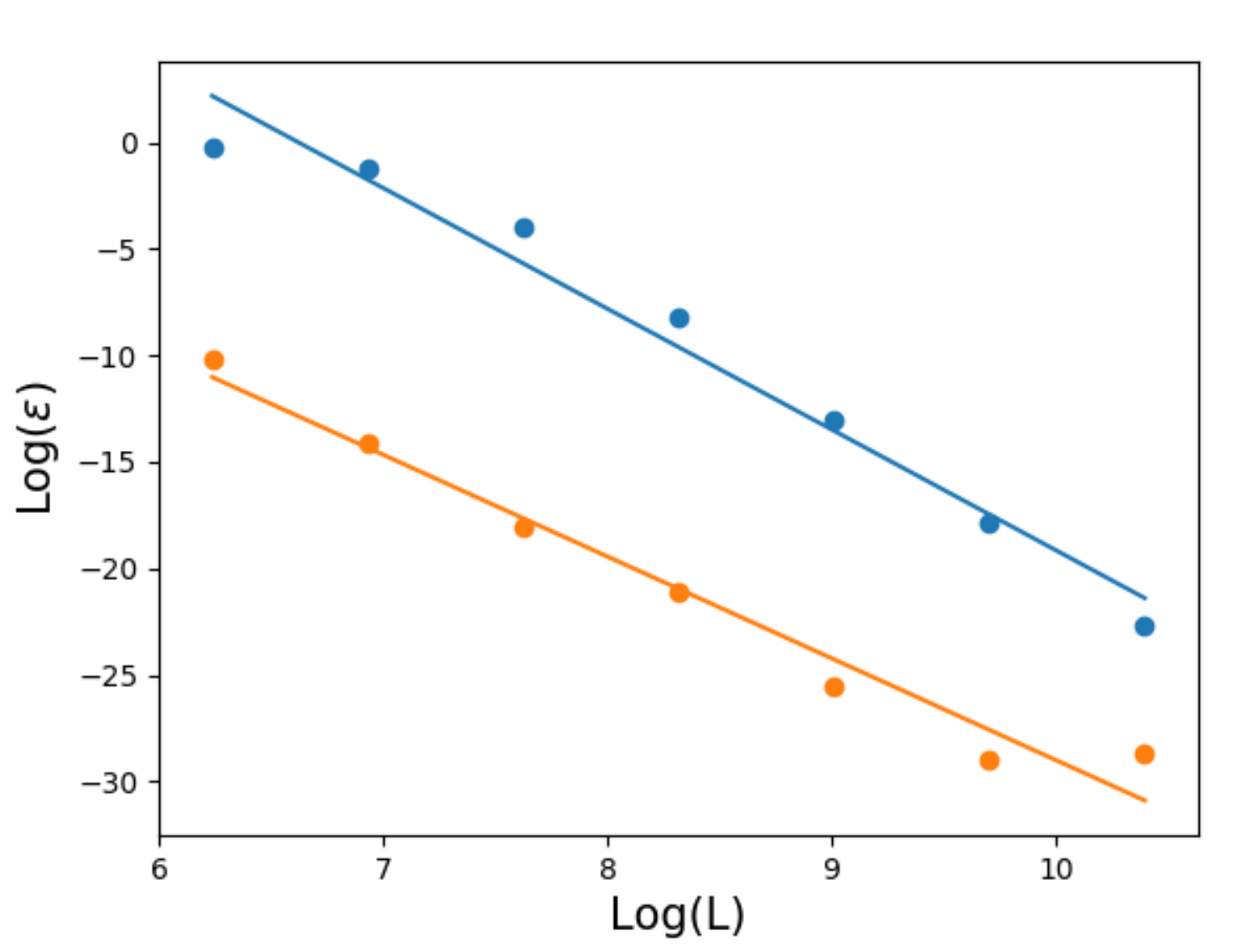} 
\caption{\footnotesize  A log-log plot of the L2 norm error $\epsilon$ as a function of lattice size $L$ for bright soliton state 14 after 1000 algorithm time steps.  The orange line is for the non-interleaved algorithm and has a slope $-4.77$.  The blue line is for the interleaved algorithm and has a slope $-5.67$.}\label{fig:L2}
\end{center}
\end{figure}

\begin{figure}[!h!t!b!p]
\begin{center}
\hspace{-.1cm}\includegraphics[width=.8\linewidth]{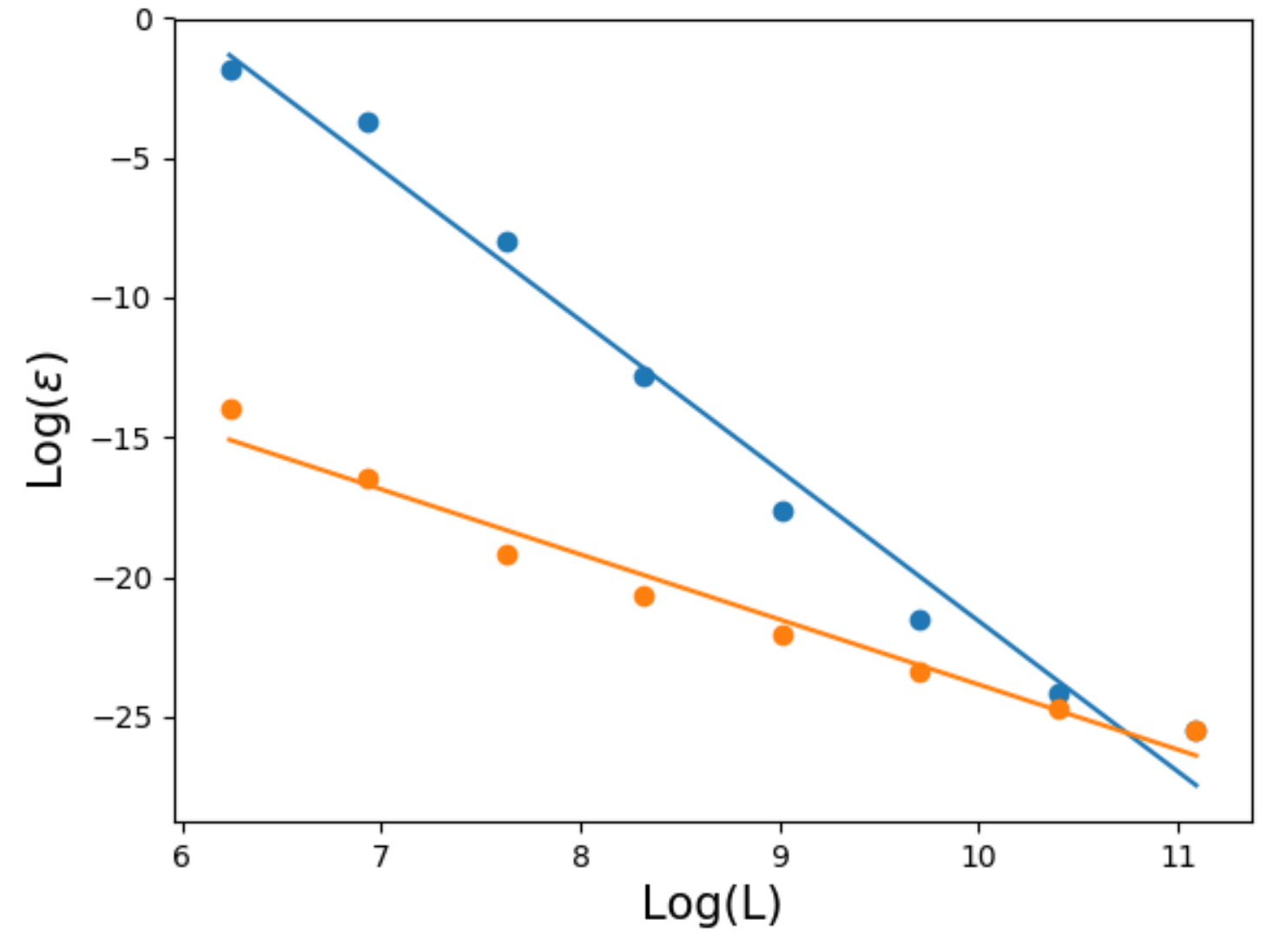} 
\caption{\footnotesize  A log-log plot of the L2 norm error $\epsilon$ as a function of lattice size $L$ for bright soliton state 14 after 1000 algorithm time steps.  The orange line is for the non-interleaved algorithm and has a slope $-2.34$.  The blue line is for the interleaved algorithm and has a slope $-5.38$.}\label{fig:L2_kink}
\end{center}
\end{figure}

So how much more accurate is the interleaved algorithm? To answer this we measure the numerical accuracy of the spin-2 BEC quantum lattice gas algorithm by calculating the L2 norm for various lattice sizes.  The idea is that because the soliton solutions are energy eigenstates of the form $\psi(x, t) = \psi(x)e^{-{i\mu t}/{\hbar}}$, with real valued $\psi(x)$, the quantity $\psi\psi^\dagger$ should remain constant as the quantum field evolves in time. Thus we measure the numerical error of the quantum lattice gas algorithm by measuring L2 norm, $\epsilon$, is given by 
\begin{align}
\epsilon = \sum_{lattice}\left(\psi\psi^{\dagger} - \psi_{sim}\psi_{sim}^{\dagger}\right)^2,
\end{align}
where the $\psi_{sim}$ is the value of the quantum field after a fixed number of iterations of the quantum lattice gas algorithm.  As the number of lattice points grows we expect the L2 norm to decrease, and the rate at which it decreases is a good measure of the scalability of a lattice gas algorithm.  It is important to note that running the algorithm on a lattice with $x$ points for a time $t$ iterations is equivalent to running the algorithm on a lattice with $\kappa x$ points for a time $\kappa^2 t$ iterations.  This is due to the diffusive ordering of space and time.  Hence, if we run the algorithm for the same number of iterations we would like to see the L2 norm decrease at a rate that is greater than $\epsilon \propto L^{-2}$ since that rate of convergence could be completely explained by the effective shorter evolution time on the larger lattice. The criteria for engineering level precision requires $\epsilon \propto L^{p}$ with $p < -4$. Figures \ref{fig:L2} and \ref{fig:L2_kink} show the rate of convergence for bright and dark solitons respectively.

As Figs.~\ref{fig:L2} and \ref{fig:L2_kink} show, the interleaved algorithm is only really crucial if there is a non-zero background field.  In the case of bright solitons with a zero background field Fig.~\ref{fig:L2} shows that both the interleaved and non interleaved algorithm meet the criteria for engineering precision since the logarithmic plot implies power laws where $\epsilon \propto L^{p}$ with $p < -4$.  However for the dark soliton state Fig.~\ref{fig:L2} shows that the while interleaved algorithm still meets the engineering level of precision with an implied power law of $p = -5.38$ the non-interleaved algorithm with an implied power law of $p = -2.34$ is barely superior to the guaranteed $\epsilon \propto L^{-2}$ from the relative scaling of space and time coming from diffusive ordering. 

\subsection{Speed, size and scalability}
We implemented the quantum lattice gas algorithm on general-purpose GPUs since they offer scalability and are inherently massively parallel computing architecture.  The current machine which runs on four NVIDIA Tesla K80 GPUs allows us to run the spin-2 BEC algorithm on up to 73 million sites in one dimension, which converts into a square grid of 8,545 sites on each side, or a three dimensional cube of 512 sites.  The GPU implementation naturally scales with the number of GPUs so a computer such as the Titan supercomputer at Oak Ridge National Laboratory could run a truly massive simulation.
\paragraph*{}
Another advantage of the GPU implementation is that the parallel nature of the GPU architecture is that it provides a massive speedup compared to traditional CPU architecture.  Table~\ref{tab:GPU speed} shows the site update rates of various implementations of the quantum lattice gas algorithm.
\begin{table}[!h!t!b!p]
\begin{center}
\begin{tabular}{|c|c|c|c|}
\hline
\multicolumn{4}{|c|}{Parallel performance} \\
\hline
Device & Processor & Language & Site Update Rate \\
\hline
& & &  \\ 
3.4 GHz Intel i5
&CPU
&Cython
& \textasciitilde 29,000 \\[.25cm]
\hline
& & &  \\ 
Nvidia 775M
&GPU
&CUDA
&\textasciitilde 190,000\\[.25cm]
\hline
& & &  \\ 
Nvidia Tesla K80
&GPU
&CUDA
& \textasciitilde 66,000,000\\[.25cm]
\hline
\end{tabular}
\end{center}
\caption{\footnotesize  The parallel performance measured in terms of the site update rates of different implementations of the quantum lattice gas algorithm. }
\label{tab:GPU speed}
\end{table}

\section{Conclusion}
\label{Section_Conclusion}

\subsection{Analytical solutions}

We carried out an investigation of possible analytical topological soliton solutions of the spin-2 GP equations in 1+1 and 2+1 spacetime dimensions. 
A summary of the analytical solutions we found for the spin-2 BEC spinor superfluid are enumerated here:
\begin{enumerate}
  \item 
  multichannel Thomas-fermi (flat) energy eigenstates;
  \item 
  multichannel bright and dark soliton energy eigenstates in 1+1 spacetime dimensions;
  \item
  a parabolic dispersion relation for bright and dark solitons 1+1 spacetime dimensions; and
  \item 
  multichannel dark soliton Pad\'e approximant solutions in 2+1 spacetime dimensions.
\end{enumerate}

\subsection{Numerical quantum simulation results}

We carried out a numerical investigation of the dynamical behavior of a spin-2 BEC spinor superfluid governed by the spin-2 GP equations in 1+1 and 2+1 spacetime dimensions. 
We compared the time evolution of the manifestly unitary spin-2 quantum lattice gas algorithm to the analytic time evolution of the state and have found that the algorithm matches theory with great numerical accuracy in both one and two dimensions.  This allows us to have confidence in the quantum lattice gas algorithm's ability to correctly predict the nonlinear physics of systems governed by the spin-2 GP equations.  The accuracy of the algorithm allows us to probe the nature of the spinor superfluid phase of a spin-2 BECs in an idealized setting where we are not hindered by the practical considerations that come with creating, maintaining, and manipulating BECs in the laboratory. 

A summary of our main numerical findings are enumerated here:
\begin{enumerate}
  \item 
  a rapidly converging L2 norm for both bright and dark soliton energy eigenstates;
  \item 
  a conservation of winding number in each $m_f$ level;
  \item 
  an agreement between the numerically calculated dispersion relation and the theoretically predicted dispersion relation;
  \item
  creation of solitons and breathers across all $m_f$ levels in soliton collisions in 1+1 spacetime dimensions; and
  \item
  pairwise creation and annihilation of vortices in soliton collisions in 2+1 spacetime dimensions.
\end{enumerate}

The real power of this quantum lattice gas algorithm is in its ability to guide our search for naturally occurring features in a spin-2 BEC. Already, the quantum simulations suggested that we should be looking for bright soliton solutions with no winding number, as well as multichannel dark vortex solutions with varying winding numbers---both of which we subsequently found and reported herein.  Since spin-2 BECs support non-Abelian interactions it is quite possible that the simulations will guide us towards an even better understanding non-Abelian phenomena in spin-2 superfluids.  

\subsection{Future outlooks}

Future experiments we would like to consider are three-dimensional quantum simulations, quantum simulations with an external trapping potential, and vortex creation experiments where the quantum simulation employ the identical techniques used to create quantum vortices in the laboratory experiment.  Each of these simulations would bring us closer to having a perfect simulation of laboratory conditions, which will accelerate our understanding of spin-2 BEC's.   
 Furthermore, three-dimensional quantum simulations of the spin-2 BEC spinor superfluid will allow us to study quantum turbulence driven strictly by unitary physics. 
 
There exists the intriguing possibility of matching a digital quantum simulation to an experimental analog quantum simulation of a spin-2 superfluid. The quantum simulation method presented here (implemented on a sufficiently large parallel array of GPUs) can be engineered to be fast enough to match (in real-time) the time-dependent behavior that can be produced and observed in a table-top spin-2 BEC experiments and spinor Fermi condensate experiments.  Running digital and analog quantum simulations in tandem (in a tightly matched way) would allows for more efficient calibration and troubleshooting of the experimental apparatus as well as provide a way to understand the results of the experimental observation of the mutual interaction of complex quantum vortices  made through free expansion of the spin-2 BEC and subsequent high-numerical aperture contrast imaging.

Finally, the spin-2 BEC Hamiltonian for a spinor superfluid is equivalent to the Hamiltonian $d$-wave superconductor. Hence,  the spin-2 quantum lattice gas method can be applied to future studies of $d$-wave superconductors such as the high-temperature YBCO superconductor.

\subsection{Acknowledgments}

JY would like to thank Professor George Vahala for discussions of  spinor GP equations and the Thomas-Fermi approximation.  All the quantum simulations reported herein were carried out in the Quantum Computing Lab at the University of Hawaii at Manoa. This basic research was supported by the grant  ``Quantum Computational Mathematics for Efficient Computational Physics" from the Computational Mathematics Program of the Air Force Office of Scientific Research.

\end{document}